\documentclass[11pt]{article}
\usepackage{mathrsfs,amssymb,amstext,amsmath,amsthm,eucal}
\usepackage{graphicx,color}
\usepackage{subfigure}
\usepackage{geometry}
\usepackage{authblk}

\newcommand{\RR}{\mathbb{R}}

\newcommand{\AdS}{\textrm{AdS}}

\DeclareMathOperator{\sech}{{sech}}
\def\identity{{\mathchoice {\rm 1\mskip-4mu l} {\rm 1\mskip-4mu l}
          {\rm 1\mskip-4.5mu l} {\rm 1\mskip-5mu l}}}
\newcommand{\lie}{\mathcal{L}}

\newcommand{\sL}{\mathfrak{sl}}

\newcommand{\SL}{\mathrm{SL}}
\newcommand{\order}{\mathscr{O}}

\usepackage{hyperref}
\hypersetup{%
  pdftitle   = {From accelerating and Poincar\texorpdfstring{\'e}{e} coordinates to black holes in spacelike warped  \texorpdfstring{$\mathrm{AdS}_3$}{AdS3}, and back},
  pdfkeywords = {3d gravity, TMG, topologically massive gravity, causal singularities, 3d black holes, warped AdS},
  pdfauthor  = {Frederic Jugeau, George Moutsopoulos, Patricia Ritter},
  pdfcreator = {\LaTeX\ with package hyperref}
}

\author{Frederic Jugeau\footnote{jugeau@ihep.ac.cn}}
\affil{
{Institute of High Energy Physics, Chinese Academy of Sciences, 
China}
}
\affil{
Theoretical Physics Center for Science Facilities, Chinese
Academy of Sciences, 
China}
\author{George Moutsopoulos\footnote{gmoutso@googlemail.com}}
\affil{
{Department of Physics 
and 
State Key Laboratory of Nuclear Physics and Technology 
Peking University, China}
}
\author{Patricia Ritter\footnote{p.d.ritter@sms.ed.ac.uk}}
\affil{
{Maxwell Institute and School of Mathematics,
University of Edinburgh, UK}}

\title{From accelerating and Poincar\'e coordinates to black holes in spacelike warped AdS$_3$, and back}

\date{\today}

\begin{document}

\maketitle

\begin{abstract}
 We first review spacelike stretched warped AdS$_3$ and we describe its black hole quotients by using accelerating and Poincar\'e coordinates. We then describe the maximal analytic extension of the black holes and present their causal diagrams. Finally, we calculate spacetime limits of the black hole phase space $(T_R,T_L)$. This is done by requiring that the identification vector $\partial_\theta$ has a finite non-zero limit. The limits we obtain are the self-dual solution in accelerating or Poincar\'e coordinates, depending respectively on whether the limiting spacetimes are non-extremal or extremal, and warped AdS$_3$ with a periodic proper time identification.
\end{abstract}

\newpage
\tableofcontents
\newpage

\section{Introduction}

A promising approach to describe the microscopic degrees of freedom in gravity lies in the conjectured AdS/CFT correspondence. The correspondence asserts that quantum gravity in its classical limit is dual to a lower-dimensional CFT in its strong regime. One can test the validity of the correspondence in different scenarios. Of particular interest is gravity in three dimensions because there the theory simplifies considerably. Pure and cosmological Einstein gravity in particular are trivial in the bulk and the solutions of the latter are described solely by their global properties~\cite{Banados:1992gq,Banados:1992wn}. Topological massive gravity (TMG) is a three-dimensional third-order gravity theory that expanded to linear order contains a single massive mode~\cite{Deser:1981wh,Deser:1982vy,renormalTMG-Deser,renormalTMG-Oda}. TMG presents a next-to-simplest model to explore quantum gravity. 

The action of TMG contains an Einstein-Hilbert term with negative cosmological constant $-1/\ell^2$ plus a gravitational Chern-Simons term
\[ 16\pi G\,S[g]=\int d^3x \sqrt{-g}\left(R + \frac{2}{\ell^2} \right)
 + \frac{\ell}{6\nu} \int d^3 x \sqrt{-g} \epsilon^{\lambda \mu \nu} \Gamma^\rho_{\lambda\sigma}
\left( \partial_\mu \Gamma^\sigma_{\rho\nu} + \frac{2}{3}\Gamma^\sigma_{\mu\tau}\Gamma^\tau_{\nu\rho}\right)~.
\]
In three dimensions, the gravitational constant $G$ has dimension of length and $\nu$ is a dimensionless positive constant that we shall take $\nu>1$. The equations of motion are
\[ R_{\mu\nu} -\frac{1}{2}R g_{\mu\nu} - \frac{1}{\ell^2} g_{\mu\nu}= +\frac{\ell}{3\nu}\epsilon_\mu{}^{\rho\sigma}(R_{\nu\rho}-\frac{1}{4}g_{\nu\rho}R)_{;\sigma}\equiv-\frac{\ell}{3\nu} C_{\mu\nu} ~,\] 
where the Cotton tensor $C_{\mu\nu}$ is a measure of conformal flatness. A solution of TMG is given by a metric along with a preferred orientation of the Levi-Civita tensor $\epsilon_{\mu\nu\rho}$. However, its solution space is more relevant to four-dimensional physics than what one might expect from such a simplification. The near-horizon geometry of the extremal Kerr black hole \cite{Bardeen:1999px}, at fixed polar angle, is a particular solution of TMG, the self-dual warped $\AdS_3$ space in Poincar\'e coordinates; see also \cite{Clement:2001ny}. The geometry of warped $\AdS_3$ plays a pivotal role here.

The last couple of years have seen a flurry of activity in TMG, due to the conjecture that the black hole solutions obtained by quotients of spacelike warped $\AdS_3$ are dual to a CFT with separate left and right central charges~\cite{Andy08}. More recently, real-time correlators were obtained for the self-dual geometry in accelerating coordinates that were chiral~\cite{Chen:2010qm}. This motivates us to take a tour of the relevant geometries, starting from the quotient construction and arriving at the self-dual warped AdS$_3$ as a spacetime limit of the black holes.

We begin in section \ref{sec:geometry} by describing the warped $\AdS_3$ geometry in three coordinate systems: the (global) warped $\AdS_3$ coordinates, accelerating coordinates and Poincar\'e coordinates. The use of the last two greatly simplifies the quotient construction of the black hole metric in section \ref{sec:quotients}.

Although the quotient construction is well-known from \cite{Andy08}, we pay particular attention to the case when causal singularities do exist behind the Killing horizons. As customary, it is for this case that we speak of a 3d black hole~\cite{Banados:1992wn}. We show that the phase space is such that the ratio of left to right temperatures $T_L/T_R$ has a lower bound, and there is a critical value of the ratio when the inner horizon coincides with the causal singularity. We also point out a subtlety in the parametrization with respect to the radii $(r_+,r_-)$ of \cite{Andy08}, namely that two separate regions of the ratio $r_+/r_-$ are in one-to-one isometric correspondence.

In section \ref{sec:Causal} we pause to describe the causal structure of the black holes. Following the previous discussion about the ratio $T_L/T_R$, we accordingly find that the causal diagrams fall into three different classes. These are similar to those of the non-extremal charged Reissner-Nordstr\"om 4d black hole (RN) for a generic ratio $T_L/T_R$, the extremal RN when $T_R=0$, and the uncharged RN when the ratio is at its critical value.

Finally, in section \ref{sec:Limits} we describe the various spacetime limits that one can take in the black hole phase space. We describe the regular extremal limit, the near-horizon limit of the extremal black holes, a near-extremal limit $T_R\rightarrow0$ for the non-extremal black holes, and the limit when both temperatures $T_R$ and $T_L$ go to zero while keeping the Hawking temperature fixed. The near-horizon and near-extremal limits give the self-dual warped $\AdS_3$ geometry in accelerating or Poincar\'e coordinates, with which we were familiarized in section \ref{sec:geometry}. The limit when both temperatures go to zero while keeping the Hawking temperature fixed gives the vacuum solution and is universal for all ratios $T_L/T_R$. The analysis is a systematic procedure and no other limits are obtained. We conclude in section \ref{sec:Discussion} with a summary and comments on our results.

The appendices contain material that is not needed in the development of the text but may complement it. Appendix A describes the hyperbolic fibration of $\mathrm{SL}(2,\RR)$, which is only briefly mentioned in \S\ref{sec:geometry}. In Appendix B we give explicit diffeomorphisms between the coordinate systems of section \ref{sec:geometry}. Although the explicit diffeomorphisms are not used in proving any of the main text's results, we present them for completeness. In appendix C we translate the bound on the ratio $T_L/T_R$ to an inequality on the ADT mass and angular momentum and the parameters of \cite{Compere:2009zj}. 

\section{Spacelike warped \texorpdfstring{$\AdS_3$}{AdS}}\label{sec:geometry}

In this section we review the geometry of spacelike
warped $\AdS_3$.  This will prepare us for a clear understanding
of the quotient construction in section \ref{sec:quotients}. We describe the metric in warped, accelerating and Poincar\'e coordinates. In summary, the metric will be written in the form
\begin{equation}\label{eq:metricf}
 g_{\ell,\nu}=\frac{\ell^2}{\nu^2+3}(-f(x) d\tau^2+\frac{dx^2}{f(x)}+\frac{4\nu^2}{\nu^2+3}(d u+x d\tau)^2),
\end{equation}
where 
\begin{equation*}
 f(x)=\begin{cases}
       x^2+1 & \text{for warped coordinates,}\\
       x^2-1 & \text{for accelerating coordinates,}\\
       x^2 & \text{for Poincar\'e coordinates.}~
      \end{cases}~
\end{equation*}
The metric \eqref{eq:metricf} satisfies the TMG equations of motion with $\epsilon_{\tau x u}=+\sqrt{-g}$. We will use the same labels $(\tau,x,u)$ for accelerating and Poincar\'e coordinates, hoping this will not cause confusion. For the warped coordinates we will use instead the coordinate labels $(\tilde{t},\sigma,\tilde{u})$, where we replace $x\rightarrow\sinh\sigma $, $u\rightarrow\tilde{u}$ and $\tau\rightarrow\tilde{t}$.

\subsection{Warped coordinates}
Let us start by expressing $\AdS_3$ as the universal cover of the special linear group
$\mathrm{SL}(2,\mathbb{R})$:
\begin{equation*}
\text{SL}(2,\mathbb{R})=\left\{  \begin{pmatrix}  T_1+X_1 &T_2+X_2 \\ X_2-T_2&T_1-X_1
                \end{pmatrix}: \quad T_1^2+T_2^2-X_1^2-X_2^2=1\right\}~.
\end{equation*}
As a group, $\text{SL}(2,\RR)$ acts on the left and right of the group manifold. We write the action as $\text{SL}(2,\RR)_L\times \text{SL}(2,\RR)_R$. We choose a basis of the right- and left-invariant 
vector fields, respectively, $l_a$ and $r_a$:
\begin{equation}\label{eq:tauvectors}
\begin{aligned}
l_1\,(r_2) &= \frac{1}{2}\left(-X_2\frac{\partial}{\partial{T_1}}-T_1\frac{\partial}{\partial{X_2}}\pm  T_2\frac{\partial}{\partial{X_1}}\pm X_1\frac{\partial}{\partial{T_2}} \right) 
\\ &
= \frac{1}{2}\begin{pmatrix} 0&-1\\-1&0 \end{pmatrix}\in\sL(2,\RR)_{L(R)}\\
l_0\,(r_0) &= \frac{1}{2}\left( -T_1\frac{\partial}{\partial{T_2}}+T_2\frac{\partial}{\partial{T_1}}\pm X_1\frac{\partial}{\partial{X_2}}\mp X_2 \frac{\partial}{\partial{X_1}} \right) 
\\ &
= \frac{1}{2}\begin{pmatrix} 0&-1\\+1&0 \end{pmatrix}\in\sL(2,\RR)_{L(R)}\\
l_2\,(r_1) &= \frac{1}{2}\left( -X_1\frac{\partial}{\partial{T_1}}-T_1\frac{\partial}{\partial{X_1}}\mp X_2\frac{\partial}{\partial{T_2}}\mp T_2\frac{\partial}{\partial{X_2}} \right) 
\\ &
= \frac{1}{2}\begin{pmatrix} -1&0\\0&+1 \end{pmatrix}\in\sL(2,\RR)_{L(R)}~.
\end{aligned}
\end{equation}
The non-zero commutators of the generators are
$[l_a,l_b]=\epsilon_{ab}^{\phantom{ab}c}l_c$ and
$[r_a,r_b]=\epsilon_{ab}^{\phantom{ab}c}r_c$, where the indices
$a=0,1,2$ are raised with a mostly-plus Lorentzian signature metric and
$\epsilon_{012}=+1$. We associate to the bases $l_a$ and $r_a$ the
dual left- and right-invariant one forms $\theta^a$ and
$\bar\theta^a$, so that $\theta^a(l_b)=\delta^a_b$ and
$\bar\theta^a(r_b)=\delta^a_b$. The Lie derivative therefore acts as
$\lie_{l_a}\theta^b=\epsilon_{a\phantom b c}^{\phantom a
  b}\theta^c$ and $\lie_{r_a}\bar\theta^b=\epsilon_{a\phantom b c}^{\phantom a
  b}\bar\theta^c$. The left-invariant one-forms allow us to write metrics on
$\text{SL}(2,\RR)$ with symmetry of rank 3,4 and 6.

Let us introduce the parametrization
\begin{equation}\label{eq:paramTTXX}
\left.\begin{array}{ll}
T_1 & = \cosh \frac{\sigma}{2} \cosh  \frac{{\tilde{u}}}{2} \cos \frac{\tilde{t}}{2}+ \sinh \frac{\sigma}{2}\sinh  \frac{{\tilde{u}}}{2} \sin \frac{\tilde{t}}{2}\\ \\
T_2 &= \cosh \frac{\sigma}{2}\cosh  \frac{{\tilde{u}}}{2} \sin \frac{\tilde{t}}{2}-\sinh \frac{\sigma}{2}\sinh  \frac{{\tilde{u}}}{2} \cos \frac{\tilde{t}}{2}\\ \\
X_1 &= \cosh \frac{\sigma}{2}\sinh  \frac{{\tilde{u}}}{2} \cos \frac{\tilde{t}}{2}+ \sinh \frac{\sigma}{2}\cosh  \frac{{\tilde{u}}}{2} \sin \frac{\tilde{t}}{2}\\ \\
X_2 &= \cosh \frac{\sigma}{2}\sinh  \frac{{\tilde{u}}}{2} \sin \frac{\tilde{t}}{2}-\sinh \frac{\sigma}{2}\cosh  \frac{{\tilde{u}}}{2} \cos \frac{\tilde{t}}{2},
\end{array}\right.
\end{equation}
which was shown in \cite{Coussaert:1994tu} to cover the whole of  $\SL(2,\RR)$ with $\tilde{u},\sigma\in\RR$ and $\tilde{t}\sim\tilde{t}+{4\pi}$. These are the hyperbolic asymmetric coordinates of \cite{Detournay:2005fz}. With the above parametrization the $\theta^a$ are
\begin{align}
 \theta^0&=-d{\tilde{t}} \cosh{\tilde{u}} \cosh\sigma +d\sigma \sinh{\tilde{u}},\\
 \theta^1&=-d\sigma \cosh{\tilde{u}} +d{\tilde{t}}\cosh\sigma\sinh{\tilde{u}},\\
 \theta^2&=d{\tilde{u}}+d{\tilde{t}}\sinh\sigma~,\\\intertext{the left-invariant vectors are}
 r_0 &=-\partial_{\tilde{t}}~,\label{eq:r0isT}\\
 r_1 &=\sin {\tilde{t}}\,\partial_\sigma
      +\cos {\tilde{t}}\tanh \sigma \,\partial_{\tilde{t}}
      +\cos {\tilde{t}}\sech\sigma\,\partial_{\tilde{u}}~,\\
 r_2 &=-\cos {\tilde{t}} \,\partial_\sigma
       +\sin {\tilde{t}} \tanh \sigma \,\partial_{\tilde{t}}
       +\sech\sigma\sin {\tilde{t}}\,\partial_{\tilde{u}}~\\
\intertext{and the right-invariant vectors are}
l_0 &= -\sinh{\tilde{u}}\,\partial_\sigma 
       -\cosh{\tilde{u}}\sech\sigma\,\partial_{\tilde{t}}
       +\cosh{\tilde{u}}\tanh\sigma \,\partial_{\tilde{u}}~,\\
l_1 &= -\cosh{\tilde{u}}\,\partial_{\sigma}
       -\sech\sigma \sinh{\tilde{u}}\,\partial_{\tilde{t}}
       +\sinh{\tilde{u}}\tanh\sigma\,\partial_{\tilde{u}}~,\\
l_2 &=\partial_{\tilde{u}}\label{eq:l2istildeu}~.
\end{align}

The Killing form, or ``round'' metric,
\[ g_{\ell}=
\frac{\ell^2}{4}\left(-\theta^0\otimes\theta^0 +\theta^1\otimes\theta^1+\theta^2\otimes\theta^2\right)~,\]
in the warped coordinates \eqref{eq:paramTTXX} becomes
\begin{equation}\label{eq:AdS3fibredoverAdS2}
g_{\ell}=\frac{\ell^2}{4}\left[-\cosh^2\sigma d{\tilde{t}}^2+d\sigma^2+(d{\tilde{u}}+\sinh\sigma d{\tilde{t}})^2\right]~.
\end{equation}
The isometry group of $\mathrm{SL}(2,\RR)$ with the round metric is $\mathrm{SO}(2,2)=(\text{SL}(2,\RR)_L \times  \text{SL}(2,\RR)_R) / \mathbb{Z}_2$, where we take into account that $-\identity$ acts similarly on each side. Unwrapping $\tilde{t}\in\RR$ gives the $\AdS_3$ metric in warped coordinates \cite{Coussaert:1994tu}. The isometry group becomes a universal diagonal cover of $(\text{SL}(2,\RR)_L\times \text{SL}(2,\RR)_R)/\mathbb{Z}_2$.

Along these lines we approach the spacelike warped metric
\begin{equation}\label{eq:wAdS3theta}
  g_{\ell,\nu}=\frac{\ell^2}{\nu^2+3}\left(-\theta^0\otimes\theta^0 +\theta^1\otimes\theta^1+\frac{4\nu^2}{\nu^2+3}\theta^2\otimes\theta^2\right)~,
\end{equation}
so that for $\nu>1$ or $\nu<1$ we have a respectively stretching or
squashing of the fiber in the direction of $l_2$ \cite{Israel:2004vv,Detournay:2005fz,Rooman:1998xf}. The isometry group is broken to that generated by the $l_2 $ and the $r_a$. In the warped coordinates $(\tilde{t},\sigma,\tilde{u})$, the warped metric is 
\begin{equation}\label{eq:wads3sigmau}
g_{\ell,\nu}=\frac{\ell^2}{\nu^2+3}\left(-\cosh^2\sigma \, d{\tilde{t}}^2+d\sigma^2+\frac{4\nu^2}{\nu^2+3}\left(d{\tilde{u}}+\sinh\sigma\, d{\tilde{t}}\right)^2\right)~.
\end{equation}
The metric is compatible with a double cover over a quadric base space, a point we elaborate on in appendix A. As before, we unwrap the time coordinate to run over $\tilde{t}\in\RR$. This is the metric of warped $\AdS_3$  in (the global) warped coordinates, which was given in \eqref{eq:metricf} for $f(x)=x^2+1$. The isometry group is the universal cover $\widetilde{\mathrm{SL}(2,\mathbb{R})}\times\RR$. 

If we compactify spacelike warped $\AdS_3$ along $l_2$, that is ${\tilde{u}}\sim {\tilde{u}}+2\pi\alpha$, we obtain the so-called self-dual solution of TMG:
\begin{equation*}
g_{\ell,\nu,\alpha}=\frac{\ell^2}{\nu^2+3}\left(-\cosh^2\sigma\, d{\tilde{t}}^2+d\sigma^2+\frac{4\nu^2}{\nu^2+3}\left(\alpha \, d\tilde{\phi}+\sinh\sigma\, d{\tilde{t}}\right)^2\right)~, 
\end{equation*}
with $\tilde{t},\sigma\in\RR$ and $\tilde{\phi}\sim\tilde{\phi}+2\pi$. The isometry group of the self-dual geometry becomes $\widetilde{\text{SL}(2,\RR)}\times U(1)$.

\subsection{Accelerating coordinates}
Let us ask how we would write the warped $\AdS_3$ metric in a
coordinate system $(\tau, x,u)$ where $\partial_\tau$ is a linear
combination of the $r_a$ and $l_2$. Since $l_2$ acts freely we can
choose $u$ to be such that $\partial_u=l_2$. The translations in $\tau$ and $u$ will be manifest symmetries of the metric.  We still need
to make an appropriate choice for the coordinate $x$, which should be
invariant under $\partial_\tau$ and $\partial_u$. That is, we require
$\partial_\tau x = \partial_u x =0$. We choose
$x=\tfrac{(\nu^2+3)^2}{4\nu^2
  \ell^2}g_{\ell,\nu}(\partial_u,\partial_\tau)$, which is indeed
invariant because $\partial_u$ and $\partial_\tau$ are Killing vectors
and they commute. The coordinate system $(\tau, x,u)$ is thus
described by the surfaces $(u,\tau)$ generated by the flows of two
Killing vectors, and a coordinate $x$ which smoothly labels them. 

Under an $\text{SL}(2,\RR)_R$ rotation on the $r_a$ and an
$\mathrm{GL}(2,\mathbb{R})$ transformation on $(u,\tau)$ we can bring
$\partial_\tau$ to one of the following forms: $r_0$, $-r_2$, or $r_0\pm r_2$. We also keep $\partial_u=l_2$ as before. The case $\partial_\tau=-r_0$ corresponds to the warped coordinates, see \eqref{eq:r0isT} and \eqref{eq:l2istildeu}. In this subsection we consider the second case, $\partial_\tau=r_2$, and in the next subsection we will consider the third case. We thus have a set of coordinates defined by the action of the Killing vectors $r_2$ and $l_2$ and their metric product. Using the metric in \eqref{eq:wads3sigmau} and the present data\footnote{that is, $\partial_\tau=r_2$, $\partial_u=l_2$ and $x=\tfrac{(\nu^2+3)^2}{4\nu^2
  \ell^2}g_{\ell,\nu}(\partial_u,\partial_\tau)=\cosh\sigma\sin \tilde t$.}, we can write the metric
\begin{equation}\label{eq:metricb}
 g_{\ell,\nu}=\frac{\ell^2}{\nu^2+3}\left(-(x^2-1) d\tau^2+\frac{dx^2}{x^2-1}+\frac{4\nu^2}{\nu^2+3}\left(du+x\, d\tau\right)^2\right)~,
\end{equation}
where we fixed $dx$ to be orthogonal to the $(u,\tau)$ hypersurfaces. 
This is precisely the metric \eqref{eq:metricf}, with $f(x)=x^2-1$. We have derived the metric form without an explicit diffeomorphism between these coordinates and warped coordinates. An explicit diffeomorphism can be found in appendix B, any other being related to it by the symmetries of the metric.

We call this set of coordinates accelerating. Accelerating coordinates have a lot in common with those of the Rindler spacetime. They describe observers with proper velocity $v=\tfrac{\partial_\tau}{|\partial_\tau|}$, whose acceleration $\nabla_v v$ is position dependent. In contrast to Rindler coordinates though, where $\partial_\tau$ is a Lorentz boost in Minkowski space, here $\partial_\tau$ is never timelike with respect to the metric. Nevertheless the $\tau$-constant surfaces are spacelike.  If we compactify along $l_2$, that is replace  $u=\alpha\phi$ in \eqref{eq:metricb}  with $\phi\sim\phi+2\pi$, we obtain the self-dual solution in accelerating coordinates: 
\begin{equation}\label{eq:SDinAccel}
 g_{\ell,\nu,\alpha}=\frac{\ell^2}{\nu^2+3}\left(-(x^2-1) d\tau^2+\frac{dx^2}{x^2-1}+\frac{4\nu^2}{\nu^2+3}\left(\alpha\,d\phi+x\, d\tau\right)^2\right)~.
\end{equation}

\begin{figure}
\begin{center}
\subfigure[integral curves of $r_2$ in warped $\AdS_3$]{\includegraphics[width=72mm]{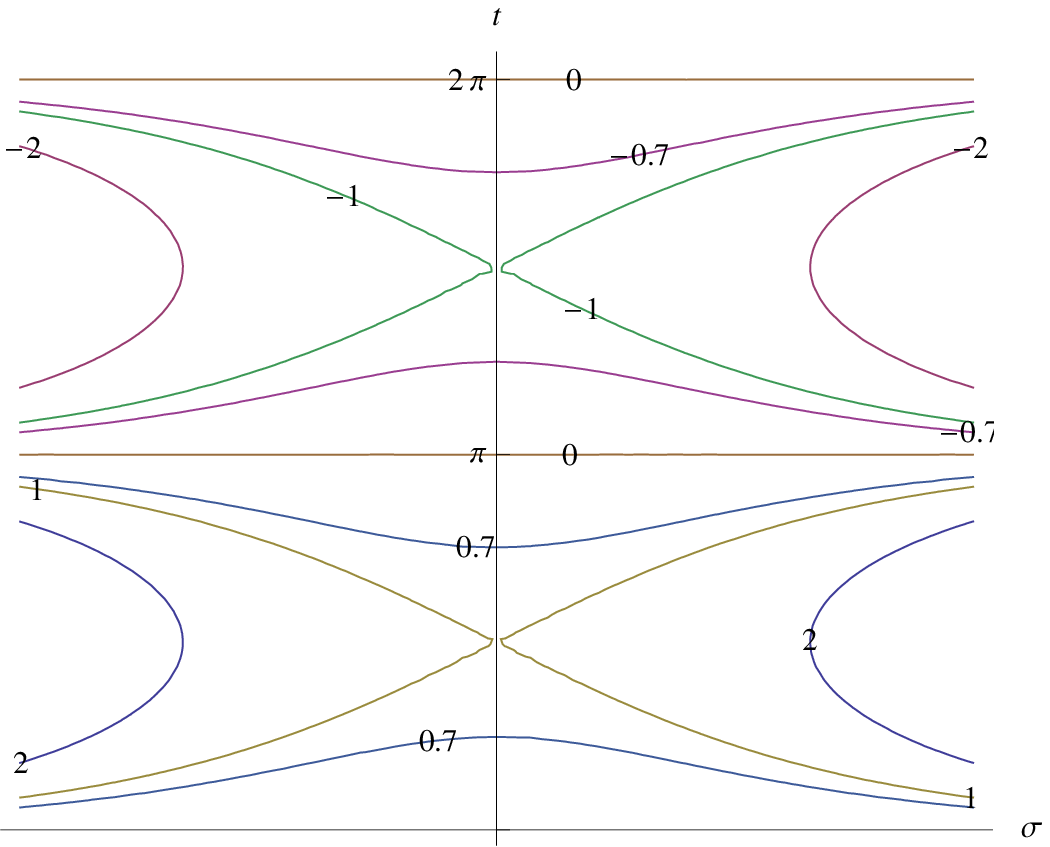}\label{fig:r2FlowInAdS3}}
\quad
\subfigure[patches of warped $\AdS_3$ covered by accelerating coordinates]{\includegraphics[width=72mm]{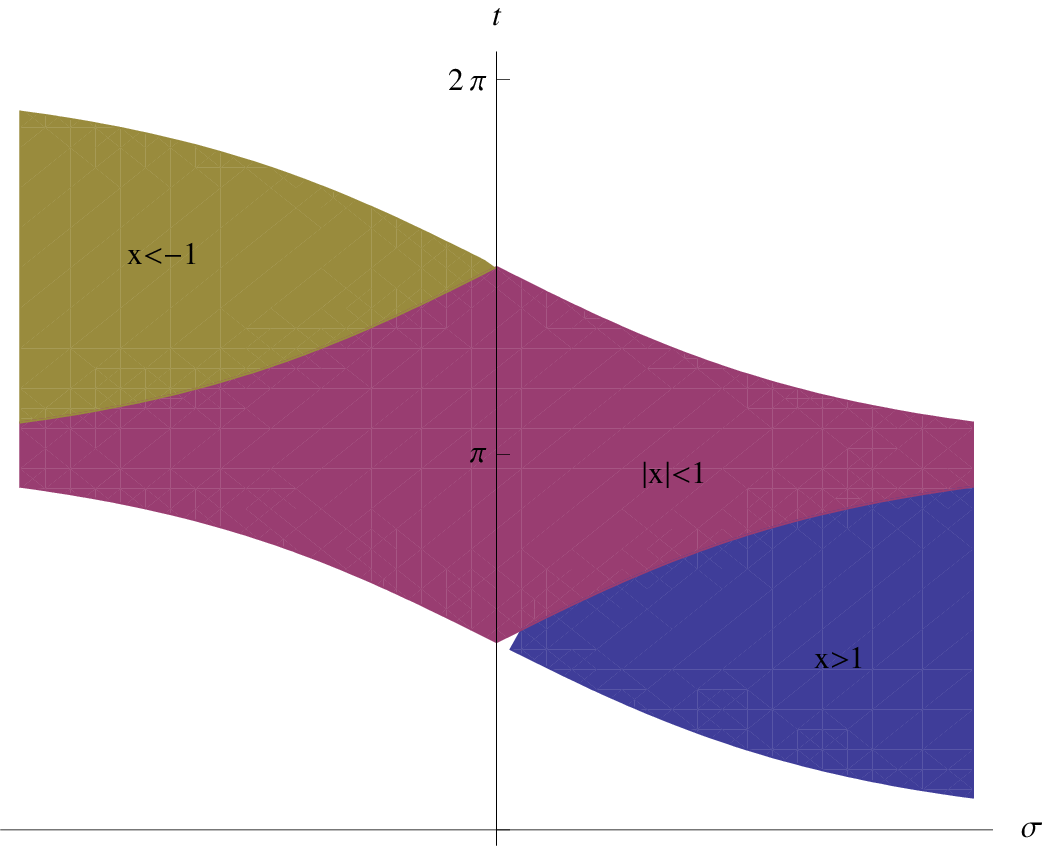}\label{fig:IsoRegions}}
\caption{The $(\sigma,\tilde t)$ plane of warped $\AdS_3$ at fixed $\tilde u$. Each line is the flow of $\partial_\tau$ and the level numbers are $x=\cosh\sigma\sin \tilde t$. At $\sigma=0$, $\tilde t=\frac{\pi}{2}\mod \pi$ we have a fixed point $r_2=0$.}
\end{center}
\end{figure}

As expected for Rindler-like coordinates, there are apparent Killing horizons appearing at $x=\pm1$. On the Killing horizons the flow of $r_2$ takes us to a line where $r_2$ becomes collinear to $l_2$. Thus the coordinates are valid only away from the Killing horizon. The warped AdS$_3$ spacetime has an infinite number of such regions. The figure in \ref{fig:r2FlowInAdS3} gives a visualisation of the situation\footnote{in the figure we take $\tilde{u}=$const., which is possible because $\tilde{u}$ is defined globally.}. The value of the level $x$ tells us where we are with respect to the Killing horizons in each region, for each of which there is an appropriate isometric embedding of $(\tau,x, u)$ in warped $\AdS_3$. Figure \ref{fig:IsoRegions} is precisely a choice of patches that the accelerating coordinates will cover.

\subsection{Poincar\'e coordinates}\label{sec:PoincCoord}

\begin{figure}
\begin{center}
\includegraphics[width=72mm]{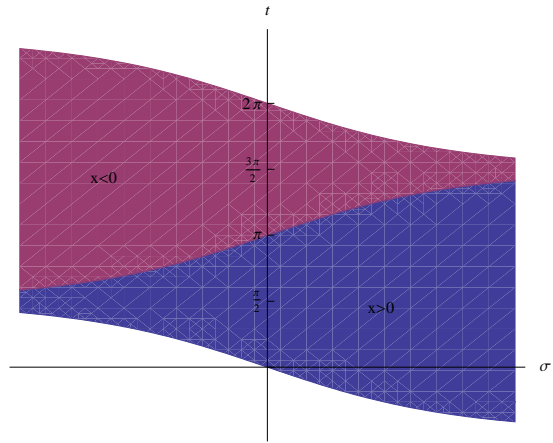}
\caption{Isometric embedding of the Poincar\'e patch.}\label{fig:PoincWedge}
\end{center}
\end{figure}

We can go through the same construction as above, but this time choosing $\partial_\tau=-r_0+r_2$. We define as before $\partial_u=l_2$ and $x=\tfrac{(\nu^2+3)^2}{4\nu^2 \ell^2}g_{\ell,\nu}(\partial_u,\partial_\tau)$. We also use the freedom to make $x$  hypersurface orthogonal. The metric is
\begin{equation}\label{eq:PoincareMetric}
g_{\ell,\nu}=\frac{\ell^2}{\nu^2+3}\left(-x^2 d\tau^2+\frac{dx^2}{x^2}+\frac{4\nu^2}{\nu^2+3}\left(du+x\, d\tau\right)^2\right),
\end{equation}
in what have been called Poincar\'e coordinates of warped $\AdS$ for obvious reasons. This is the metric in \eqref{eq:metricf} with $f(x)=x^2$. Similarly to the accelerating coordinates, the surface $x=0$ is a Killing horizon. Figure \ref{fig:PoincWedge} shows a Poincar\'e patch of warped AdS$_3$. An explicit diffeomorphism can be found in Appendix B.

The case $\partial_\tau=r_0+r_2$ is similar to the above, simply by the warped AdS$_3$ discreet symmetry $(\tilde{t},\tilde{u})\mapsto(-\tilde{t},-\tilde{u})$ that flips the sign of $r_0$ while preserving that of $r_2$.  Compactifying along $l_2$, that is $u\sim u+2\pi\alpha$, gives us the self-dual solution in Poincar\'e coordinates
\begin{equation}\label{eq:SDinPoinc}
g_{\ell,\nu,\alpha}=\frac{\ell^2}{\nu^2+3}\left(-x^2 d\tau^2+\frac{dx^2}{x^2}+\frac{4\nu^2}{\nu^2+3}\left(\alpha \,d\phi+x\, d\tau\right)^2\right).
\end{equation}

\section{Black hole quotients}\label{sec:quotients}

Here we will follow the construction of \cite{Andy08}, and find the quotients of spacelike warped AdS$_3$ that have causal singularities hidden behind Killing horizons. We quotient spacelike warped AdS$_3$ by $\exp(2\pi\partial_\theta)$ with $\partial_\theta$ given by
\begin{equation}\label{eq:quotientthetaTLTR}
 \partial_\theta = \begin{cases} 2\pi\ell \,T_R \, r_2 + 2 \pi\ell \, T_L \, l_2 &\text{non-extremal black holes}\\  (r_2\pm r_0) + 2 \pi\ell \, T_L \, l_2 & \text{extremal black holes}.\end{cases}
\end{equation}
The timelike case $\partial_\theta = A\, r_0 + B\, l_2$ yields naked closed timelike curves (CTCs)  and so we do not consider it. Up to an $\mathrm{SL}(2,\RR)_R$ rotation, which is an isometry of warped AdS, these three cases cover all choices of  $\partial_\theta$. The quotient by \eqref{eq:quotientthetaTLTR} defines the left and right temperatures $T_L$ and $T_R$ in analogy to the BTZ case.

We pay attention to two points of interest. The first is that singular
regions of a non-extremal quotient can be hidden behind a Killing
horizon only when $T_L/T_R$ is bigger than a critical value. The
second is that the Ansatz for $T_L$ and $T_R$ as a function of $r_+$
and $r_-$ in \cite{Andy08} is not one-to-one for $T_L/T_R$ smaller than a second (different) critical value.

The method we employ is to describe the quotient in accelerating or,
for the case of extremal black holes, Poincar\'e coordinates. The
reason is quite simple: other than $\partial_\theta$ we would like a
metric where the translation $\partial_t$ is the remaining and manifest isometry. The
coordinates $(t,\theta)$ should then be given  by a $\mathrm{GL}(2,\mathbb{R})$ transformation on the accelerating,
respectively Poincar\'e, coordinates $(\tau,u)$, see \eqref{eq:quotientthetaTLTR}. The remaining radial
coordinate $r$ is then any function of $x$ that labels the integral
flows of $(\partial_\tau,\partial_u)$. The non-extremal black hole
horizons are none other than the Killing horizons of warped AdS$_3$ at
$x=\pm 1$, while the extremal black hole horizon lies on the Poincar\'e horizon $x=0$.

\subsection{Non-extremal black holes}\label{sec:nonextrquot}
Assume the accelerating coordinates $(\tau,x,u)$ and the quotient defined by
\begin{equation}\label{eq:quotmatrixGen}
 \begin{pmatrix}
  t\\\theta
 \end{pmatrix}=
\begin{pmatrix}
 a &b\\c&d
\end{pmatrix}
\begin{pmatrix}
 \tau\\u
\end{pmatrix}~.
\end{equation}
The periodicity $\theta\sim\theta+2\pi$ is preserved under the coordinate transformation
\begin{equation}\label{eq:GaugeofT}
 \begin{pmatrix}
  t'\\\theta'
 \end{pmatrix}=
\begin{pmatrix}
 A&0\\B&1
\end{pmatrix}_{A\neq0}
\begin{pmatrix}
  t\\\theta
 \end{pmatrix}.
\end{equation}
That is, the quotient matrix in \eqref{eq:quotmatrixGen} is equivalent under
\begin{equation*}
 \begin{pmatrix}
  a&b\\c&d
 \end{pmatrix}\approx
\begin{pmatrix}
 A a& Ab\\a B+c&Bb+d
\end{pmatrix}~.
\end{equation*}

When $b=0$ we bring the matrix to the form
\begin{equation}\label{eq:SDMatrixAccel}
 \begin{pmatrix}
  t\\\theta
 \end{pmatrix}=
\begin{pmatrix}
 1 &0\\0&\frac{1}{\alpha}
\end{pmatrix}
\begin{pmatrix}
 \tau\\u
\end{pmatrix}~.
\end{equation}
This quotient is the self-dual solution, albeit in accelerating coordinates, where the parameter $\alpha$ is the one in \eqref{eq:SDinAccel}. When $b\neq0$ we bring the matrix to the form
\begin{align}
  \begin{pmatrix}
  t\\\theta
 \end{pmatrix}&=
\frac{2\nu}{\nu^2+3}\begin{pmatrix}
a&1\\c&0
\end{pmatrix}
\begin{pmatrix}
 \tau\\u
\end{pmatrix}~\label{eq:BTZquotient}
\\
\Leftrightarrow
  \begin{pmatrix}
  \tau\\u
 \end{pmatrix}&=
\frac{\nu^2+3}{2\nu}\begin{pmatrix}
0&1/c\\1&-a/c
\end{pmatrix}
\begin{pmatrix}
 t\\\theta
\end{pmatrix}~
\\
\Leftrightarrow
\label{eq:KVquotientMatrix}
\begin{pmatrix}
  \partial_t\\\partial_\theta
 \end{pmatrix}&=
\frac{\nu^2+3}{2\nu}\begin{pmatrix}
0&1\\1/c&-a/c
\end{pmatrix}
\begin{pmatrix}
 \partial_\tau\\\partial_u
\end{pmatrix}~,
\end{align}
where our choice is to normalize the length $|\partial_t|^2=\ell^2$. In this case, we identify 
\begin{align}
2\pi\ell\, T_R &= \frac{\nu^2+3}{2\nu} \frac{1}{c} \\
2\pi\ell\, T_L &= - \frac{\nu^2+3}{2\nu} \frac{a}{c} ~.
\end{align}
By reflecting $\theta\mapsto -\theta$ if necessary, we choose $c>0$. Note that $1/c\neq0$ and so we cannot describe the extremal case $T_R= 0$ regularly.

\begin{figure}
 \begin{center}
  \input{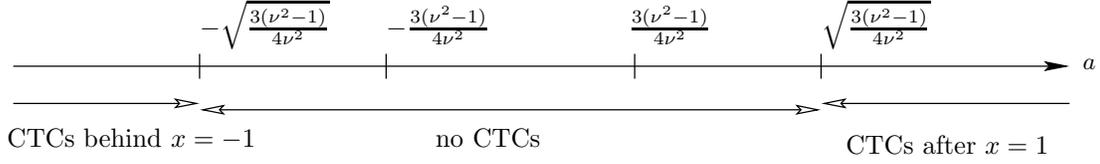}\caption{CTCs versus the parameter $a$.}\label{fig:CTCVSa}
 \end{center}
\end{figure}

We now ask when singular regions $|\partial_\theta|^2\leq0$ exist and whether they are hidden behind the Killing horizon $x=1$. Observe that we have not yet restricted the parameter $a=-T_L/T_R$ in \eqref{eq:BTZquotient}. A simple calculation in accelerating coordinates reveals
\begin{align*}
 c^2|\partial_\theta|^2&=\ell^2\frac{\nu^2+3}{4\nu^2}\left(-(x^2-1)+\frac{4\nu^2}{\nu^2+3} (x-a)^2\right),\\\intertext{with determinant}
 \Delta_x &= \ell^4 \frac{\nu^2+3}{\nu^2}\left(a^2-3\frac{\nu^2-1}{4\nu^2}\right)\\\intertext{and}
 \partial_x (c^2 |\partial_\theta|^2)&=\ell^2\frac{\nu^2+3}{2\nu^2}\left(3\frac{\nu^2-1}{\nu^2+3}x-\frac{4\nu^2}{\nu^2+3}a\right)~.
\end{align*}
It follows that for $|a|<\frac{\sqrt{3(\nu^2-1)}}{2\nu}$ there are no CTCs, for $a<-\frac{\sqrt{3(\nu^2-1)}}{2\nu}$ CTCs exist in $x<-1$ and for $a>\frac{\sqrt{3(\nu^2-1)}}{2\nu}$ there are CTCs after $x>1$. This is summarized in figure \ref{fig:CTCVSa}. In fact, the values $a>\frac{\sqrt{3(\nu^2-1)}}{2\nu}$ tell us that $x>1$ is an accelerating patch where CTCs exist. One can then move by the discreet isometry $(x,u)\mapsto(-x,-u)$ to the outer region of the black hole. This essentially flips the sign of $a$, so that the ratio $\tfrac{T_L}{T_R}$ is bounded from below by $\frac{\sqrt{3(\nu^2-1)}}{2\nu}$.

The quotient in \cite{Andy08} is parametrized by $(r_+,r_-)$, where the right and left temperatures are related to $T_R$ and $T_L$ by
\begin{align}
 \label{eq:TR}
 T_R&=\frac{(\nu^2+3)(r_+-r_-)}{8\pi\ell},\\
 \label{eq:TL}
 T_L&=\frac{\nu^2+3}{8\pi\ell}\left(r_++r_--\frac{\sqrt{r_+ r_-({\nu^2+3})}}{\nu}\right)~,
\end{align}
and the radial coordinate is chosen to be
\begin{equation}
r=\frac{r_+-r_-}{2} x+\frac{r_++r_-}{2}\label{eq:BHrWRTx}~.
\end{equation}
The coordinate transformation from the accelerating coordinates $(\tau,x,u)$ to the black hole coordinates $(r,t,\theta)$ is given by \eqref{eq:BTZquotient} and \eqref{eq:BHrWRTx}. By using \eqref{eq:metricb}, we find the black hole metric in the ADM form
\begin{equation}\label{eq:BlackHoleADM}
 ds^2=-N^2 dt^2+\frac{\ell^4 dr^2}{4 R^2 N^2}+\ell^2 R^2 (d\theta+ N^\theta dt)^2,
\end{equation}
where
\begin{align}
 R^2&=\frac{3(\nu^2-1)}{4} r(r-r_0)\\
 N^2&= \frac{\ell^2 (\nu^2+3)}{4 R^2}(r-r_+)(r-r_-)=\frac{\ell^2(\nu^2+3)}{3(\nu^2-1)}\frac{(r-r_-)(r-r_+)}{r(r-r_0)}\\
 N_\theta&=\frac{2\nu r - \sqrt{r_+ r_- (\nu^2+3)}}{2 R^2}\\
r_0&=\frac{ 4\nu\sqrt{r_+ r_- (\nu^2+3)}-(\nu^2+3)(r_++r_-)}{3(\nu^2-1)}\label{eq:r0}~.
\end{align}
The Levi-Civita tensor transforms to $\epsilon_{tr\theta}=+\sqrt{-g}$.

\begin{figure}
 \begin{center}
  \input{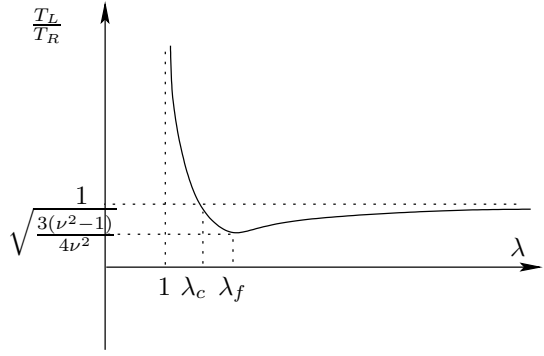}
  \caption{$\frac{T_L}{T_R}$ versus the ratio $\lambda=r_+/r_->1$.}\label{fig:aVSl}
 \end{center}
\end{figure}

Let us now point out a subtlety in the parametrization with respect to $r_+$ and $r_-$ in \eqref{eq:TR} and \eqref{eq:TL}. From these equations, it follows that the ratio of temperatures 
\begin{align}
\frac{T_L}{T_R}=\frac{\nu(r_++r_-)-\sqrt{r_+ r_-(\nu^2+3)}}{\nu(r_+-r_-)}~\label{eq:BTZquotientPar}
\end{align}
can be written as a function of the ratio $\lambda=r_+/r_->1$. At the same time, the right temperature $T_R$ can remain arbitrary positive for any given $\lambda$ by adjusting $r_->0$. It is instructive to draw the graph of the ratio $T_L/T_R$ as a function of $\lambda$, see figure \ref{fig:aVSl}. We find that $T_L/T_R$ decreases from plus infinity until the minimum at
\begin{align}\label{eq:lambdaf}
\lambda_f&=1+6\frac{\nu^2-1}{\nu^2+3}\left(1+\frac{\sqrt{3}}{3}\frac{2\nu}{\sqrt{\nu^2-1}}\right)~,\\\intertext{for which}
\left.\frac{T_L}{T_R}\right|_{\lambda_f}&=\sqrt{\frac{3(\nu^2-1)}{4\nu^2}}<1~.\notag
\end{align}
It then increases asymptotically to the value of $1$. On the one hand, this agrees with the necessary bound on $T_L/T_R$. On the other, we find that there is a hidden isometry between pairs $(r_+,r_-)$ in the two regions $(\lambda_c,\lambda_f)$ and $(\lambda_f,\infty)$, 
 where $
  \lambda_c = \frac{4\nu^2}{\nu^2+3}$ {gives} $
  \left.\frac{T_L}{T_R}\right|_{\lambda_c} = 1$. Indeed, two values of the ratio of radii, say $r_+/r_-$ and $\tilde{r}_+/\tilde{r}_-$, can give the same value of $T_L/T_R$ and so, the two pairs $(r_+,r_-)$ and $(\tilde{r}_+,\tilde{r}_-)$ will result in the same two invariants $T_L$ and $T_R$ that define the quotient in \eqref{eq:quotientthetaTLTR}.
  
More precisely, the isometry relates black hole metrics with
\begin{align}
 r_+&= \frac{\nu^2+3}{3(\nu^2-1)}\left(\sqrt{\tilde{r}_-}-\frac{2\nu}{\sqrt{\nu^2+3}}\sqrt{\tilde{r}_+}\right)^2\\
 r_-&= \frac{\nu^2+3}{3(\nu^2-1)}\left(\frac{2\nu}{\sqrt{\nu^2+3}}\sqrt{\tilde{r}_-}-\sqrt{\tilde{r}_+}\right)^2,~
\end{align}
while the radial coordinate transforms as $r\mapsto \tilde{r}$ according to
\begin{equation}
 \frac{2 r - r_+ - r_-}{r_+-r_-} = \frac{2 \tilde r - \tilde{r}_+ - \tilde{r}_-}{\tilde{r}_+-\tilde{r}_-}~.
\end{equation}
The last equation is true by its equality to the coordinate label $x=\tfrac{(\nu^2+3)^2}{4\nu^2
  \ell^2}g_{\ell,\nu}(\partial_u,\partial_\tau)$, see \eqref{eq:BHrWRTx}, which is fixed for given $T_L$ and $T_R$.

\begin{figure}
\begin{center}
\input{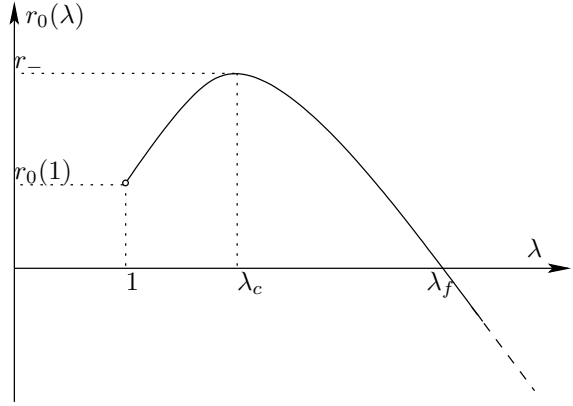}\caption{Graph of $r_0(r_+/r_-)$ for fixed $(\nu,r_-)$.}\label{fig:r0lambda}        
\end{center}
\end{figure}

Note that $r_0$ in \eqref{eq:r0}, as a function of the ratio
$\lambda\equiv r_+/r_-\geq1$ with $r_-$ fixed, presents a
maximum $r_0(\lambda_c)=r_-$ and then decreases
monotonously, as in figure \ref{fig:r0lambda}. In particular, $r_0(\lambda_f)=0$. 
As a result, the maximum root of $R(r)^2$, denoted
$\bar{r}_0$ hereafter, is
\begin{equation*}
\bar{r}_0=\left\{ \begin{array}{ll} 0&\textrm{if}\;\;r_0<0\;\;i.e.\;\;\lambda>\lambda_f\\
r_0&\textrm{if}\;\;r_0\in[0,r_-]\;\;i.e.\;\;1\leq\lambda\leq\lambda_f,
\end{array}~ \right.
\end{equation*}
and so $R(r)^2>0$ for $r>r_-$ and the CTCs are always hidden behind $r_-$. The equality $R(r_-)^2=0$ holds only for $r_0(\lambda_c)=r_-$, that is when the inner horizon coincides with the singularity. For later use, let us also define the shorthand $R^2_\pm$:
\begin{equation*} R^2_{\pm}\equiv
R(r_{\pm})^2=\frac{r_{\pm}}{4}\left(2\nu\sqrt{r_{\pm}}-\sqrt{(\nu^2+3)r_{\mp}}\right)^2\geq0~.
\end{equation*}

We should stress that we arrive at global results using accelerating coordinates. This is because $\partial_\theta$ in \eqref{eq:quotientthetaTLTR} is a global identification and one can choose to cover any of the infinite regions discussed in \S\ref{sec:geometry} using accelerating coordinates. The Killing horizons at $r=r_-,\,r_+$ are inherited from the accelerating horizon at $x=\pm1$. The lower bound in $T_L/T_R$ was discussed in \cite[\S6.1.1]{Andy08}. The parametrization of $T_L$ and $T_R$ in terms of $r_-$ and $r_+$ in \cite{Andy08} is such that the lower bound is satisfied. A subtle feature of the parametrization is the isometry in parameter space for $r_+/r_-\geq\lambda_c$. Let us also comment that, by the above analysis, the parameter assignment $T_L=0$ appears special and disconnected from the region $T_L/T_R>\frac{\sqrt{3(\nu^2-1)}}{2\nu}$. We will nevertheless obtain it as the vacuum limit of the non-extremal black holes in \S\ref{sec:Limits}. 

\subsection{Extremal black holes}
The quotient that gives the extremal black holes in terms of the second Killing vector in \eqref{eq:quotientthetaTLTR} does not present any particular point of interest. We can repeat the previous derivation \emph{mutatis mutandis}, where now the coordinates $(\tau,u)$ in \eqref{eq:quotmatrixGen} are the Poincar\'e coordinates of warped AdS. The case $b=0$, see \eqref{eq:SDMatrixAccel}, is the self-dual solution in Poincar\'e coordinates \eqref{eq:SDinPoinc}. The case $b\neq 0$ gives the black hole solution in ADM form \eqref{eq:BlackHoleADM} by setting $r_+=r_-$ in \eqref{eq:TL} and using $x=r-r_-$. The singular regions are behind $r<r_-$ for all values of $T_L\neq0$, which can be chosen positive by reflecting $\theta$ if necessary. 

In the non-extremal quotient given by the matrix in \eqref{eq:KVquotientMatrix}, observe that the parameter $1/c\sim T_R$ is always positive. One can thus never reach the extremal black holes $T_R=0$ from a regular quotient of that type. It is clear though that the non-extremal black holes have an extremal limit given by setting $r_+=r_-$ in the metric  \eqref{eq:BlackHoleADM}. We shall recover this result in section \ref{sec:Limits} as a limit of the non-extremal quotient \eqref{eq:KVquotientMatrix}. 

However, let us first examine the causal structure of the black holes. We will see that the critical value $r_+/r_-=\lambda_c$ is special with respect to the causal structure. We thus have three cases to consider: generic non-extremal, the case $r_+/r_-=\lambda_c$, and the extremal case.

\section{Causal structure}\label{sec:Causal}
In this section we will examine the causal structure of the spacelike warped black holes in a manner similar to \cite{Banados:1992gq}. Although these geometries are ideal, they are likely to appear as the end state of physical processes where chronology is protected. We will show that the Penrose-Carter diagram of a generic non-extremal or extremal black hole is similar to the 4d non-extremal, respectively extremal, Reissner-Nordstr\"om black hole. Recall that we uncovered a critical value $r_0=r_-$ (for $r_+/r_-=\lambda_c$)  that is isometric to $r_-=0$. We accordingly find that the $r_-=r_0$ black hole has a causal diagram similar to that of the Schwarzschild black hole, that is the uncharged Reissner-Nordstr\"om black hole. 

In what follows we will work with the two-dimensional metric $g_2$
\begin{equation*}
  g=\underbrace{-N^2 dt^2+\frac{\ell^4 dr^2}{4 R^2 N^2}}_{g_2}+\ell^2 R^2 (d\theta+ N_\theta dt)^2~.
\end{equation*}
If a curve $\gamma:[0,1]\rightarrow M$ has tangent vector $\dot{\gamma}\in \gamma^* TM$, then
\[ g_2(\dot{\gamma},\dot{\gamma}) > 0 \Longrightarrow g(\dot{\gamma},\dot{\gamma})>0 ~,\]
thus a causal curve $\gamma$ must be non-positive on $g_2$
\[ g(\dot{\gamma},\dot{\gamma})\leq 0 \Longrightarrow g_2(\dot{\gamma},\dot{\gamma})\leq0~.\]
On the other hand, any causal curve $g_2(\dot{\gamma},\dot{\gamma})\leq0$ can be lifted to a causal curve on $g$, e.g. by choosing the horizontal lift
\begin{equation}\label{eq:horizLift}\dot{\theta}+N_{\theta}\dot{t}=0~.\end{equation}
Let us note that the metric $g_2$ does not capture the behaviour of causal geodesics, see e.g. \cite{0256.53016}. However null curves on $g$ such that \eqref{eq:horizLift} holds are geodesic on $g_2$. They correspond to zero angular momentum $p_\theta=g(\dot{\gamma},\partial_\theta)$.

The metric $g_2$ then tells us about all causal relations by neglecting the angle $\theta$. One might wonder why we do not take a $\theta=\text{const.}$ section. After disentangling the angle one can indeed find a Kruskal extension, as done generically in \cite{Racz:1992bp}. However, the angle is not defined globally on the different Kruskal patches, so our choice is simpler since the connection $d\theta+N_\theta dt$ is global. Furthermore, a local $\theta$-section will not give us information on causal relations, nor can it be compatible with any geodesic. Indeed, observe that the restriction of the metric on a constant angle will always be positive definite far away from the horizon.

The similarities with the RN black holes are not coincidental. Our method involves reducing the causal properties to the two-dimensional quotient space under the angular isometry $\partial_\theta$. The difference to the Reissner-Nordstr\"om solution then, other than the dimensionality of the sphere, is a non-trivial connection one-form $d\theta+N_\theta dt$, compare e.g. with Carter's extension in \cite{Carter1966423}.

We will first describe the future horizon ingoing coordinates. This is done so as to intermediately introduce the Regge-Wheeler tortoise coordinate $r_*$. We then write down the Kruskal-Szekeres extension. 
We shall also use the ingoing coordinates in section \ref{sec:Limits}, in order to derive the near-horizon geometry of extremal black holes. 

\subsection{Ingoing Eddington-Finkelstein coordinates}
To introduce Eddington-Finkelstein coordinates, one first solves for the Regge-Wheeler tortoise coordinate $r_*$, which in our case satisfies
\begin{equation}\label{eq:tortDef}
\frac{dr_*}{dr} = \frac{\ell^2}{2 R N^2}=
\frac{\sqrt{3(\nu^2-1)}}{\nu^2+3}\frac{\sqrt{r(r-r_0)}}{(r-r_-)(r-r_+)}~.
\end{equation}
For $r>\bar{r}_0$ and $r_+\neq r_-$, the solution is branched as follows
\begin{multline}\label{eq:tortSolGen}
r_*=\frac{\sqrt{3(\nu^2-1)}}{\nu^2+3}\Bigg(
\frac{\sqrt{r_+(r_+-r_0)}}{r_+-r_-}\ln\left(\frac{|r-r_+|}{\left(\sqrt{r}\sqrt{r_+-r_0}+\sqrt{r-r_0}\sqrt{r_+}\right)^2}\right)\\
- \frac{\sqrt{r_-(r_--r_0)}}{r_+-r_-}\ln\left(\frac{|r-r_-|}{\left(\sqrt{r}\sqrt{r_--r_0}+\sqrt{r-r_0}\sqrt{r_-}\right)^2}\right)\\+2\ln\left(\sqrt{r}+\sqrt{r-r_0}\right)
\Bigg)~.
\end{multline}
For the critical value $r_+/r_-=4\nu^2/(\nu^2+3)$, the solution \eqref{eq:tortSolGen} is also well-defined. For the extremal case $r_+=r_-$, \eqref{eq:tortDef} becomes
\begin{equation}\label{eq:tortDefExt} \frac{dr_*}{dr} =\frac{ \sqrt{3(\nu^2-1)}}{\nu^2+3}\frac{\sqrt{r(r-r_0)}}{(r-r_-)^2}\end{equation}
and its solution is branched as
\begin{multline}\label{eq:tortSolExt}
  r_*=\frac{\sqrt{3(\nu^2-1)}}{\nu^2+3}\Bigg(-\frac{\sqrt{r(r-r_0)}}{r-r_-}+2\ln(\sqrt{r}+\sqrt{r-r_0})\\+\frac{1}{2}\frac{2 r_--r_0}{\sqrt{r_-(-r_0+r_-)}}\ln\frac{|r-r_-|}{(\sqrt{r(r_--r_0)}+\sqrt{r_-(r-r_0)})^2}\Bigg)~.
\end{multline}
The ingoing coordinate is defined as $u=t+r_*$.

The coordinates $(u,r)$ are well-defined on and past the future horizon. In contrast, the angle $\theta$ is entangled, that is it diverges for geodesics that cross the horizon. For $r_+\neq r_-$ and $r_+/r_-\neq4\nu^2/(\nu^2+3)$ we define the angle
\begin{multline*}
  \theta_{in}=\theta\\ - \frac{4\nu}{\nu^2+3}\frac{1}{\nu(r_+-r_-)}\left(
 -\frac{2\nu r_++\sqrt{r_+ r_-(\nu^2+3)}}{2\nu r_+-\sqrt{r_+ r_-(\nu^2+3)}}\ln\left(\sqrt{r(r_+-r_0)}+\sqrt{r_+(r-r_0)}\right)\right.\\\left.+
 \frac{2\nu r_-+\sqrt{r_+ r_-(\nu^2+3)}}{|2\nu r_--\sqrt{r_+ r_-(\nu^2+3)}|}\ln\left(\sqrt{r(r_--r_0)}+\sqrt{r_-(r-r_0)}\right)\right)+N_{\theta}(r_+)u~,
\end{multline*}
while for $r_+/r_-=4\nu^2/(\nu^2+3)$ we define 
\[ \theta_{in}=\theta -\frac{4}{r_-3(\nu^2-1)}\ln\left(\sqrt{r}+\frac{2\nu}{\sqrt{3(\nu^2-1)}}\sqrt{r-r_-}\right)+N_{\theta}(r_+)u~.\]
For the extremal black holes $r_+=r_-$ we define
\begin{multline*}
\theta_{in}=\theta+N_\theta(r_-)u + \frac{4\nu}{\sqrt{3(\nu^2-1)}(\nu^2+3)}\Bigg(
   -\frac{\sqrt{r(r-r_0)}}{r_-(r-r_-)}\\+\frac{ r_0}{2 r_-\sqrt{r_-(-r_0+r_-)}}
    \ln\frac{r-r_-}{(\sqrt{r(r_--r_0)}+\sqrt{r_-(r-r_0)})^2}\Bigg)~.
    \end{multline*}

These definitions are such that, in $(u,r,\theta_{in})$ coordinates, in all cases the metric becomes 
\begin{equation}\label{eq:GenMetricIng} 
g = -N^2 du^2+\frac{\ell^2}{R}dr du+\ell^2 R^2(d\theta_{in}+ {N}_{\theta_{in}} du)^2~,
\end{equation}
with ${N}_{\theta_{in}}(r)=N_{\theta}(r)-N_{\theta}(r_+)$ being zero on the horizon. 
The coordinates $(u,r,\theta_{in})$ are regular on the future horizon $r=r_+$ and valid until $r=r_-$. The Hamiltonian of a free-falling particle is 
\[ \mathcal{H}=\frac{2}{\ell^4}\left(\ell^2 R p_u p_r+N^2 R^2 p_r^2+\frac{\ell^2}{4 R^2} p_{\theta_{in}}^2-\ell^2 R {N}_{\theta_{in}} p_r p_{\theta_{in}}\right) \]
where $p_{\theta_{in}}$, $p_u$ are constants of motion. 
Null geodesics, $\mathcal{H}=0$, satisfy
\[ \dot{u}=\frac{2}{\ell^2} R p_r ~\]
and for $p_{\theta_{in}}=0$ the ingoing rays are those with $p_r \equiv 0$. 


\subsection{Kruskal extension of non-extremal black holes}

We first describe the Kruskal extension across $r=r_+$ for the case $r_+\neq r_-$. With
\begin{align*}
  b_+&=\frac{\nu^2+3}{4}\frac{r_+-r_-}{R_+}=\frac{1}{2}\frac{r_+-r_-}{\sqrt{r_+(r_+-r_0)}}\frac{\nu^2+3}{\sqrt{3(\nu^2-1)}} 
\end{align*}
and $\rho(r)=e^{b_+ r_*}$, define
\begin{align*}
\left.\begin{aligned}
  U&=\rho(r) e^{b_+ t}\\
 V&=\rho(r) e^{-b_+ t}\\
 \theta_+&= \theta - \frac{N_{\theta}(r_+)}{2 b_+}\ln\frac{U}{V}
\end{aligned}\right\}&\text{ for }r> r_+\quad\text{ and }\\ \\
\left.\begin{aligned}
 U&=\rho(r) e^{b_+ t}\\
 V&=-\rho(r) e^{-b_+ t}\\
 \theta_+&= \theta + \frac{N_{\theta}(r_+)}{2 b_+}\ln\frac{U}{V}~
\end{aligned}\right\}&\text{ for } r_-<r< r_+~.
\end{align*}
The transformation in $r_-<r<r_+$ is given so that one can match the Kruskal patches using \eqref{eq:BlackHoleADM}. In these coordinates, the metric becomes
\begin{equation}
 \label{eq:wBTZKrMetK1}
 ds^2 = \Omega_+^2 dU dV + \ell^2 R^2\left(d\theta_+ + N_{UV}(VdU-U dV)\right)^2~,
 \end{equation}
where
\begin{multline*}
 \Omega^2_+ =\frac{4\ell^2}{\nu^2+3}\frac{r_+(r_+-r_0)}{(r_+-r_-)^2} \frac{\left(r-r_-\right)^{1+\sqrt{\frac{r_-(r_--r_0)}{r_+(r_+-r_0)}}}}{r(r-r_0)}(\sqrt{r}\sqrt{r_+-r_0}+\sqrt{r-r_0}\sqrt{r_+})^2\\
 \times \left(\sqrt{r}\sqrt{r_--r_0}+\sqrt{r_-}\sqrt{r-r_0}\right)^{-2\sqrt{\frac{r_-(r_--r_0)}{r_+(r_+-r_0)}}} \left(\sqrt{r}+\sqrt{r-r_0}\right)^{-2\frac{r_+-r_-}{\sqrt{r_+(r_+-r_0)}}}~
\end{multline*}
is everywhere positive 
and $N_{UV}$ can be shown to be regular at $r=r_+$. The coordinate $r$ is given implicitly by $U V=\rho^2(r)$, which is monotonous in $r>r_+$ and, separately, in $r_-<r\leq r_+$. We have the limits $\lim_{r\rightarrow+\infty}UV=+\infty$, $\lim_{r\rightarrow r_+}UV=0$ and $\lim_{r\rightarrow r_-{}^+}UV=-\infty$. 
We can extend with the isometry $V\mapsto -V$ and $U\mapsto -U$, and the patch $K_+=\{U,V\in\RR\}$ is regular everywhere with a metric given by \eqref{eq:wBTZKrMetK1}. 

We now build an extension across $r_-$ for $r_+\neq r_-$ and $r_-\neq r_0$. With
\begin{align*}
  b_-&=-\frac{\nu^2+3}{4}\frac{r_+-r_-}{R_-}=-\frac{1}{2}\frac{r_+-r_-}{\sqrt{r_-(r_--r_0)}}\frac{\nu^2+3}{\sqrt{3(\nu^2-1)}}
\end{align*}
and $\rho(r)=e^{b_- r_*}~,$ define
\begin{align*}
\left.\begin{aligned}
  \tilde{U}&=\tilde\rho(r) e^{b_- t}\\
 \tilde{V}&=\tilde\rho(r) e^{-b_- t}\\
 \theta_-&=\theta - \frac{N_{\theta}(r_-)}{2 b_-}\ln\frac{\tilde{U}}{\tilde{V}}
\end{aligned}\right\}&\text{ for }\bar{r}_0<r< r_-\quad\text{ and }\\ \\
\left.\begin{aligned}
 \tilde{U}&=-\tilde\rho(r) e^{b_- t}\\
 \tilde{V}&=\tilde\rho(r) e^{-b_- t}\\
 \theta_-&=\theta + \frac{N_{\theta}(r_-)}{2 b_-}\ln\frac{\tilde{U}}{\tilde{V}}
\end{aligned}\right\}&\text{ for } r_-< r<r_+~.
\end{align*}
The metric becomes
\begin{equation}
 \label{eq:wBTZKrMetK2}
 ds^2 = \Omega_-^2 d\tilde{U} d\tilde{V} + \ell^2 R^2(d\theta_- + N_{\tilde{U}\tilde{V}}(\tilde{V}d\tilde{U}-\tilde{U} d\tilde{V}))^2
 \end{equation}
with
\begin{multline*}
 \Omega^2_- =\frac{4\ell^2}{\nu^2+3}\frac{r_-(r_--r_0)}{(r_+-r_-)^2} \frac{\left(r_+-r\right)^{1+\sqrt{\frac{r_+(r_+-r_0)}{r_-(r_--r_0)}}}}{r(r-r_0)}\left(\sqrt{r}\sqrt{r_--r_0}+\sqrt{r-r_0}\sqrt{r_-}\right)^2\\
 \times \Big(\sqrt{r}\sqrt{r_+-r_0}+\sqrt{r_+}\sqrt{r-r_0}\Big)^{-2\sqrt{\frac{r_+(r_+-r_0)}{r_-(r_--r_0)}}} \Big(\sqrt{r}+\sqrt{r-r_0}\Big)^{2\frac{r_+-r_-}{\sqrt{r_-(r_--r_0)}}}
\end{multline*}
and $r$ is given implicitly by $\tilde{U}\tilde{V}$, which is again monotonous in $r$. We have the limits $\lim_{r\rightarrow \bar{r}_0{}^+}\tilde{U}\tilde{V}=\rho_0^2>0$, $\lim_{r\rightarrow r_-}\tilde{U}\tilde{V}=0$ and $\lim_{r\rightarrow r_+{}^-}\tilde{U}\tilde{V}=-\infty$. 

We similarly extend the coordinate range with the isometry $U\mapsto -U$, $V\mapsto -V$. The patch $K_-=\{\tilde{U},\tilde{V}\in\RR\}$ is defined regularly throughout with the metric given in \eqref{eq:wBTZKrMetK2}.

\begin{figure}
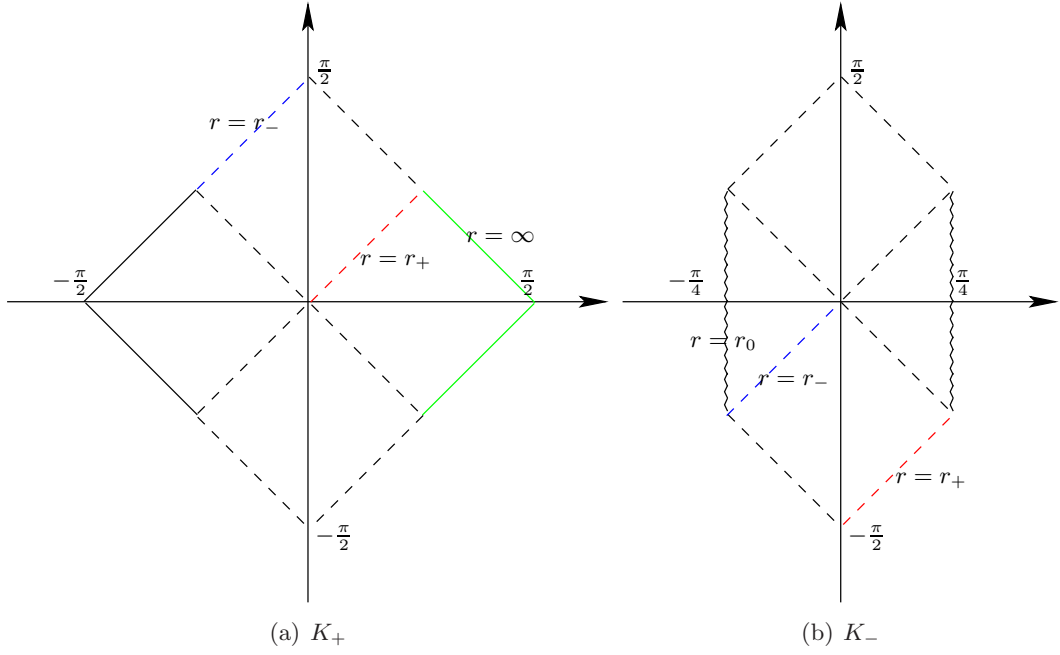

 \begin{center}
 \subfigure[{$K_+$}]{{\input{figK1reg.pspdftex}}}
 \subfigure[{$K_-$}]{{\input{figK2reg.pspdftex}}}
\caption{Penrose diagrams of Kruskal patches for $r_0\neq r_-$ black holes.}\label{fig:PenReg}
 \end{center}
\end{figure}

\begin{figure}
 \begin{center}
\input{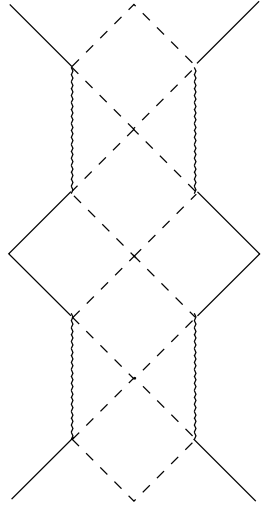}\caption{The Penrose diagram of maximally extended $r_0\neq r_-$ black holes.}\label{fig:MaxPenReg}
 \end{center}
\end{figure}

By transforming into the finite-range coordinates $\tan(u)=U$ and $\tan(v)=V$, and similarly $\tan(\tilde{u})=\rho_0\,\tilde{U}$ and $\tan(\tilde{v})=\rho_0\,\tilde{V}$, we draw in figure \ref{fig:PenReg} the Carter-Penrose diagrams for the two patches. Note that the conformal factor multiplying the connection one-form in the metric blows up as
\[ \frac{R^2}{U^2 V^2 \Omega_+^2} \sim \mathcal{O}\left(r^{\frac{r_+-r_-}{\sqrt{r_+(r_+-r_0)}}}\right)~.\]
To circumvent any ambiguity, we compactify the manifold by using instead the coordinate system
\begin{equation*}\begin{aligned}
   \hat{U}&=    U^{z(U)+1}\\
   \hat{V}&=V^{z(V)+1}~,
                \end{aligned}
                \end{equation*}
where the exponent $z(x)$ is a function that is zero for small but positive $x$ and grows smoothly within a finite range up to the constant value of $\frac{r_+-r_-}{\sqrt{r_+(r_+-r_0)}}$. The factor multiplying the connection one-form then becomes finite and non-vanishing in the limit $r\rightarrow\infty$. The maximal extension is  obtained by concutting $K_+$ after $K_-$ ad infinitum, as in figure \ref{fig:MaxPenReg}.

\begin{figure}
 \begin{center}
 {\input{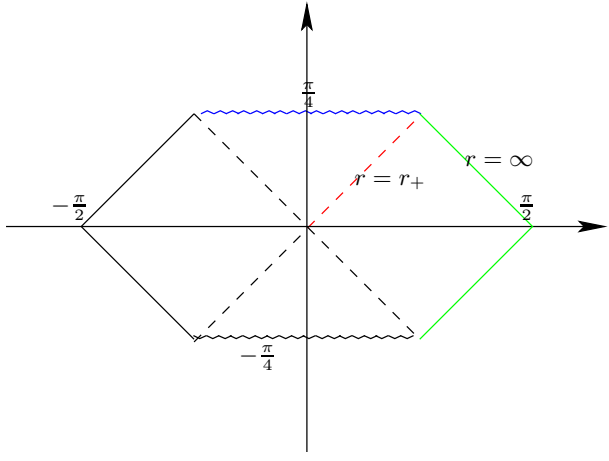}}
\caption{Penrose diagram for $r_0=r_-$.}\label{fig:PenCrit}
 \end{center}
\end{figure}

For the critical value $r_+/r_-=4\nu^2/(\nu^2+3)$ we define the patch $K_+$ as before. With the special value
\[ b_+ = \frac{\nu^2+3}{4\nu}~,\]
we find
\[ \Omega_+^2=\frac{4\ell^2}{3(\nu^2-1)}\frac{4\nu^2}{\nu^2+3}\frac{1}{r}\left(\sqrt{r}\sqrt{3(\nu^2-1)}+2\nu\sqrt{r-r_-}\right)^2\Big(\sqrt{r}+\sqrt{r-r_-}\Big)^{-\frac{\sqrt{3(\nu^2-1)}}{\nu}}~.\]
However, here we do not extend beyond the inner horizon $r_-$ where $|\partial_\theta|^2<0$. 
The Kruskal coordinates have the limits $ \lim_{r\rightarrow+\infty}UV=+\infty$, $\lim_{r\rightarrow r_+}UV=0$ and
\begin{equation*}
 \lim_{r\rightarrow 0}UV=-\rho^2_0=-\frac{1}{\nu^2+3}r_-^{\sqrt{\frac{3(\nu^2-1)}{2\nu}}}~.
\end{equation*}
The Penrose diagram of the critical black hole is drawn in \ref{fig:PenCrit}, where we use $U=\rho_0 \tan(u)$ and $V=\rho_0 \tan(v)$.

\subsection{Kruskal extension of extremal black holes}
Finally, we describe the extremal case. We present the conformal compactification at once, by using a  transformation similar to the one for the extremal Reissner-Nordstr\"om in \cite{Carter1966423}. 
However, some care is needed to show that the connection one-form is also well-defined. Using the tortoise coordinate, define for $r>r_-$
\begin{align}
 \tan U &= t+r_*\\
 \tan V &= - t + r_* \\
 \theta_{UV} &=
 \theta - N_{\theta}(r_-) t - C \left( 2\tanh^{-1}\tan\frac{U}{2} - 2\tanh^{-1}\tan\frac{V}{2}\right), \label{eq:thetaUVExt}
\end{align}
with the constant 
\begin{equation*}
 C=- \frac{4\nu}{(\nu^2+3)\sqrt{3(\nu^2-1)}\sqrt{r_-(r_--r_0)}}~.
\end{equation*}
The metric takes the form
\begin{equation*}
 g=\Omega^2 dU dV + \ell^2 R^2 (d\theta_{UV} + \tilde{N}_{UV} (dU -dV) )^2,
\end{equation*}
with
\begin{align*}
 \Omega^2&= \frac{N^2}{\cos^2 U \cos^2 V}~
\end{align*}
and $\tilde{N}_{UV}$ is zero on the horizon. We first observe that $\Omega^2$ is non-zero on the future and past horizon. Indeed, the dangerous factor $ \frac{(r-r_-)^2}{\cos^2 V}$ in the limit $V\rightarrow 0$ goes like
\begin{equation*}
\begin{array}{ll}
\left( \frac{1}{\cos V} \right) \left( \frac{1}{r-r_-} \right)^{-1} &\longrightarrow
  \left( 2\frac{\sin V}{\cos^2 V} \right) \left( - \frac{\frac{\partial r}{\partial V}}{(r-r_-)^2} \right)^{-1}\\ \\ 
\ &\longrightarrow 2 \frac{\sqrt{3(\nu^2-1)}}{\nu^2+3}\sqrt{r_-(r_--r_0)}~,
\end{array}
\end{equation*}
where the last equation uses the derivative of the tortoise coorindate in \eqref{eq:tortDefExt}. It follows that $\lim_{V\rightarrow 0} \Omega^2$ is finite and  non-vanishing on the future horizon, and similarly on the past horizon. We also defined $\theta_{UV}$ in \eqref{eq:thetaUVExt} with the term linear in $C$ so that a potential pole of $g(\partial_\theta,\partial_U)\sim \tilde{N}_{UV}$ in $r-r_-$ vanishes.

\begin{figure}
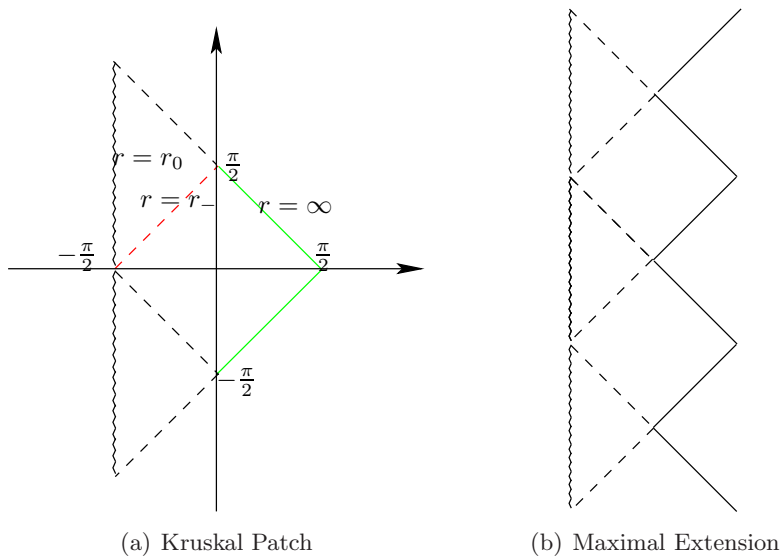

 \begin{center}
 \subfigure[{Kruskal Patch}]{{\input{figK1ext.pspdftex}}}\qquad
 \subfigure[{Maximal Extension}]{\makebox[0.3\textwidth]{\input{figK11ext.pspdftex}}}
\caption{Penrose diagrams of extremal black holes.}\label{fig:PenExt}
 \end{center}
\end{figure}

Altogether, this means that we can use the same transformation on and behind the horizon but for a different domain of $U,V$, and by replacing $C\rightarrow - C$. The singular region is at $\tan U + \tan V =2 r_*$ which can be brought to zero by a suitable shift in $r_*$. The Penrose diagram of the extremal black hole is drawn in figure \ref{fig:PenExt} and the maximal extension can be obtained with the isometry $U-V\mapsto U-V + 2\pi\mathbb{Z}$.

\section{Spacetime limits}\label{sec:Limits}

In the previous sections we explored the geometry of warped AdS, its black hole quotients and their causal properties. In particular, the extremal black holes are obtained from a different quotient than their non-extremal counterparts. At the same time, the extremal black holes are a regular limit of the non-extremal black holes, in the sense that we can set $r_-=r_+$ in the ADM form. In this section we explain this limit in more detail. 

We also want to ask what other classical\footnote{that is, we consider $\ell$, $G$ and $\nu$ fixed.} limits we can obtain from the warped AdS$_3$ black holes. We will obtain the near-horizon geometry of extremal black holes and we will define several other spacetime limits, which give us the self-dual warped AdS, in either accelerating or Poincar\'e coordinates, and warped AdS$_3$ with a proper time identification.

We find it helpful to recall Geroch's notion of a spacetime limit \cite{Geroch:1969ca}. Here one collects a family of metric spacetimes $(M_L,g_L)$, where $L>0$, and constructs the augmented manifold $\mathscr{M}=\{ (M_L,g_L,L)_{L}\}$. A spacetime limit, $L\rightarrow 0$, is \emph{invariantly} defined on the boundary of $\mathscr{M}$. Spacetime limits are interesting for the properties of the family $(M_L,g_L)$ that are inherited in the limit, a typical example being that the rank of Killing vectors and Killing spinors \cite{Blau:2002mw} does not reduce. Naturally, the spacetime limit $(M_0,g_0)$ is of interest when its maximal extension is not included in the original phase space.  

An instance of Geroch's notion is when there is a local isometry $f_L:M_L\rightarrow M_1$, for $L>0$, between the metrics $g_L$ and $g_1$. The limit can then be said to be of the metric itself $g_1$ rather than a limit in the family of metrics $g_L$. An example is the Penrose limit \cite{Blau:2002mw}. A \emph{metric limit} typically involves blowing up a neighbourhood of the spacetime.  Minkowski space is not only a spacetime limit of 4d black holes, where the mass $M\equiv L\rightarrow 0$, but can also be written in terms of a metric limit\cite{Geroch:1969ca}. In the latter case, one is translating in the limit to the asymptotically flat region while keeping the mass $M$ fixed. In this paper, we call a metric limit the near-horizon geometry of $g_1$ when the isometry $f_L$ fixes the outer horizon. 

In our case, the non-extremal black hole metrics are parametrized by $(T_R,T_L)$ which we take as functions of $L>0$. Each black hole in the phase space is given by the identification Killing vector $\partial_\theta$ as written in \eqref{eq:quotientthetaTLTR}. Note, though, that the identification vector in \eqref{eq:quotientthetaTLTR} is unique up to $\mathrm{SL}(2,\RR)_R$ rotations. The question we ask is, what are the limits of the non-extremal black holes as $T_R\rightarrow 0$.

In order to simplify our discussion, we do not ask what happens in the limit behind the outer horizon. We thus take the $M_L$ to cover only part of the maximally extended spacetime. In practice this means we can work with the accelerating, or Poincar\'e coordinates, and define the limits explicitly. The coordinates will thus depend explicitly on $L$. This description is complementary to the previous not only for practical reasons, but also because it describes the relation of the coordinate range of the limit manifold $M_0$ to that of $M_{L>0}$.

We first describe the near-horizon limit of the extremal black holes (\S\ref{sec:nhGeomCoord}), using the coordinate description in the framework of \cite{Reall:2002bh,Kunduri:2008tk,Clement:2004yr}. We then consider spacetime limits of non-extremal black holes when $T_R \rightarrow 0$. There are two such limits. The first gives us a geometry similar to the near-horizon geometry of the extremal ones, but in accelerating coordinates (\S\ref{sec:SDLimit}). We call this the near-extremal limit. The second one gives the extremal black holes (\S\ref{sec:extrlim}). We also describe the near-horizon geometry of extremal black holes in the invariant description (\S\ref{sec:nhliminv}). Finally, we consider the case when we send $T_R\rightarrow 0$ while keeping the Hawking temperature fixed (\S\ref{sec:vaclim}).

\subsection{Near-horizon limit}\label{sec:nhGeomCoord}
Let us erect Gaussian null coordinates on the future horizon of a spacelike warped black hole, as explained in \cite{Friedrich:1998wq}. The ingoing coordinates $(u,r,\theta_{in})$ are such that $\theta_{in}$ is a well-defined angle on a spacelike section of the horizon and $u$ is the group parameter of $\xi=\partial_u$. Recall that the metric in ingoing coordinates has the form \eqref{eq:GenMetricIng}:
\[ g = -N^2 du^2+\frac{\ell^2}{R}dr du+\ell^2 R^2(d\theta_{in}+ {N}_{\theta_{in}} du)^2~.\]
We are interested in defining a new coordinate $\bar{r}$ that is the affine parameter of a null geodesic congruence $\gamma$ emanating from the horizon and parametrised by $(u,\theta_{in})$. We fix its velocity $\dot{\gamma}_0$ on the future horizon $\mathscr{H}^+$ to be the normalized null complement of $\partial_u$ and $\partial_{\theta_{in}}$ with respect to the metric: $g(\dot{\gamma}_0,\partial_u)|_{\mathscr{H}^+}=1/2$ and $g(\dot{\gamma}_0,\partial_{\theta_{in}})|_{\mathscr{H}^+}=0$. The Hamiltonian of a free-falling particle and its geodesic equations are
\begin{align*}
 \mathcal{H}&=\frac{2}{\ell^4}\left(\ell^2 R p_u p_r+N^2 R^2 p_r^2+\frac{\ell^2}{4 R^2} p_{\theta_{in}}^2-\ell^2 R {N}_{\theta_{in}} p_r p_{\theta_{in}}\right)\\
 \dot{r}&=\frac{2}{\ell^4}\left(\ell^2 R p_u + 2 N^2 R^2 p_r\right)\\
 \dot{\theta}&=\frac{2}{\ell^2}\left(\frac{p_{\theta_{in}}}{2 R^2}- R {N}_{\theta_{in}} p_r\right)\\
 \dot{u}&=\frac{2}{\ell^2}R p_r~.
\end{align*}
The equations can easily be solved. The constraint $\mathcal{H}=0$ implies $p_r|_{\mathscr{H}^+}=p_{\theta_{in}}=0$ and with $p_u=\frac{1}{2}$ we find $p_r\equiv0$, $\dot{\theta}=\dot{u}=0$ and
\begin{equation}\label{eq:affineGauss} \frac{dr}{d\bar{r}} = \frac{R}{\ell^2} ~,\end{equation}
where $\bar{r}$ is the affine parameter. This equation is solved generically by
\begin{align}\label{eq:affineGaussSol}
 r&=r_0\cosh^2\left(\frac{\sqrt{3(\nu^2-1)}}{4\ell^2}\bar{r}-c\right),\\\intertext{where}
 \cosh c&= \sqrt\frac{r_+}{r_0}\text{ and }c>0~.
\end{align}
The coordinate transformation \eqref{eq:affineGaussSol} covers the region $r\in(r_0,+\infty)$, which corresponds to $$\bar{r}\in\left(-\infty,\frac{4\ell^2}{\sqrt{3(\nu^2-1)}}c\right).$$ The other coordinates remain $u\in\mathbb{R}$ and $\theta_{in}$ periodic.

For $r_+\neq r_-$ the metric takes the form
\begin{equation}\label{eq:nonExtGauss} g = -\bar{r}\, F(\bar{r})\, du^2+d\bar{r} du+\ell^2 R^2(r(\bar{r}))(d\theta_{in}+ N_{\theta_{in}}(r(\bar{r}))\, du)^2~,\end{equation}
where $N^2=\bar{r}\,F(\bar{r})$ and $F(\bar{r})$ is regular non-vanishing on the horizon $\bar{r}=0$. It follows that the near-horizon limit \emph{cannot} be defined for non-extremal black-holes. Indeed, if we assume a diffeomorphism $\bar{r}\mapsto \bar{r}/L$ that zooms in on a neighbourhood of the horizon, then the component $g_{u{\bar{r}}}$ dictates an appropriate rescaling $u\mapsto L u $ so that $\text{lim}_{L\rightarrow 0}g_{u{\bar{r}}}$ remains finite. However, this blows up the component $g_{uu}$.

When $r_+=r_-$, $F(\bar{r})=\bar{r} \, H(\bar{r})$ where $H(\bar{r})$ is regular non-vanishing at $\bar{r}=0$. Introducing the coordinate transformation
\begin{equation}\begin{aligned}\label{eq:nhgeomCoordL}
 \bar{r}'&= \bar{r}/L\\
 u'&= L u,
\end{aligned}\end{equation}
and sending $L\rightarrow 0$, gives the metric limit
\begin{equation}\label{eq:nhgeomCoordMetric}
 g= \frac{\nu^2+3}{4\ell^2} \bar{r}'^2\,du'^2 + d\bar{r}' du' + \ell^2 R^2_-\left(d\theta_{in}+\left.\frac{d N_{\theta_{in}}}{dr}\right|_{r_-}\frac{R_-}{\ell^2} \bar{r}' du' \right)^2~,
\end{equation}
with 
\[  \left.\frac{d N_{\theta_{in}}}{dr}\right|_{r_-} = \frac{4}{r_-^2}\frac{\nu-2\nu^2 r_- r_-+\nu\sqrt{\nu^2+3} r_-}{(2\nu-\sqrt{\nu^2+3})^2}~.\]
Observe that as $L\rightarrow0$, any point $\bar{r}$ close to $\bar{r}=0$ is pushed away to infinity with respect to $\bar{r}'$. The metric in \eqref{eq:nhgeomCoordMetric} is the self-dual solution with $\alpha =\frac{\nu^2+3}{2\nu}R_-$ in Poincar\'e coordinates. This can be verified by using the diffeomorphism
\begin{equation}\label{eq:nhLimitIsPoincDiff}\begin{aligned}
  u'&=\tau-\frac{1}{x}\\
 \bar{r}'&=\frac{2\ell^2}{\nu^2+3}x\\
 \phi&=\theta + \frac{2\nu}{\nu^2+3}\frac{1}{R_-}\ln x~.
\end{aligned}\end{equation}

The above derivation zooms indefinitely into the future horizon of an extremal black hole along a geodesic congruence. Using the coordinate description we got the self-dual warped AdS$_3$ in Poincar\'e coordinates. This result is universal. We would not have been able to arrive at the same geometry in, say, accelerating or warped coordinates. Since the horizon is non-bifurcate the same should be true for the limit spacetime. One could use equivalently the double null coordinates $(u,v)$, where $u$ is the ingoing and $v=-t+r_*$ is the outgoing coordinate. The description using $(u,v)$ serves to show that we are zooming in on the whole of the horizon. Finally, we could have used the ADM coordinates $(r,t)$. The limit is given by $r-r_-=r'\,L$ and $t=  t'/L$. This description provides an equivalent explanation for why the limit is in Poincar\'e coordinates. This is the case because $t$ is defined asymptotically by observers who wish to probe the horizon. As such, the near-horizon inherits a preferred time which is not related to the global warped time $\tilde{t}$.

We can already ask what properties are inherited in the limit. It is clear that one such property is the nature of the horizon. The size of the radius of $\theta$ on the horizon is also inherited, this being a consequence of definition \eqref{eq:nhgeomCoordL} as an isometry that fixes the horizon. We will later describe the near-horizon geometry invariantly, using the identification vector $\partial_\theta$, and see that this is related to the extremal black hole via $\alpha=2\pi\ell\, T_L$.

\subsection{Near-extremal limit}\label{sec:SDLimit}
Although a non-extremal black hole does not admit a near-horizon limit, we can consider a limit in the black hole phase space $(T_L,T_R)$ for $T_R\rightarrow 0$. This limit cannot be considered a metric limit because $T_R$ is continuously varied. Furthermore, there is more than one way to take the limit. Here we will consider the case when the limit gives us the self-dual solution in accelerating coordinates. We call the limit the near-extremal near-horizon limit, or near-extremal limit for short, and we stress it is a spacetime limit in the phase space of non-extremal black holes.

A black hole is described by $(T_R,T_L)$ that enter the definition \eqref{eq:quotientthetaTLTR} of the Killing vector $\partial_\theta$,
\[ \partial_\theta =  2\pi\ell\,T_R \, r_2  + 2\pi\ell\,T_L \,l_2~.\]
 There are however \emph{two} gauge freedoms that we can use in its description. The first is an active $\text{SL}(2,R)_R$ rotation that isometrically maps the outer region as embedded in warped AdS$_3$ to a new region. The rotation transforms $r_2 \mapsto A\, r_2 + B\, r_0$, with $A^2-B^2=1$, and we can use instead the vector
\begin{equation}\label{eq:SL2RonIKVne}
 \partial_{\theta'} = 2\pi\ell\,T_R \left(  A \, r_2 \pm B \, r_0\right) + 2\pi\ell\,T_L\, l_2~.
\end{equation}
Note that we are considering an active transformation in warped $\AdS$. That is, the rotation $\exp(\tanh^{-1}(\frac{B}{A})\, r_1)$ is not an isometry of the metric. 

The second gauge freedom is how we describe time $t$. The $\mathrm{GL}(2,\RR)$ diffeomorphism in \eqref{eq:GaugeofT} keeps the identification vector $\partial_\theta$ invariant. However, we are redefining $\partial_t$ and so the metric form  in the new coordinate system does change. It is this freedom that we shall use and fix here. Indeed, notice that if we simply take $T_R=0$ in \eqref{eq:KVquotientMatrix}, that is send $1/c\rightarrow 0$ and keep $a/c$ fixed in \eqref{eq:KVquotientMatrix}, we end up with $\partial_t$ collinear with $\partial_\theta$. The coordinates $(t,\theta)$ are thus ill-defined in the limit. We use the transformation
\begin{equation}\label{eq:selflimitAB}
 \begin{pmatrix}
  t'\\\theta'
 \end{pmatrix}=
\begin{pmatrix}
 -\frac{1}{b}\frac{\nu^2+3}{2\nu}\frac{T_R}{T_L} &0\\\frac{\nu^2+3}{2\nu}\frac{1}{2\pi \ell \, T_L}&1
\end{pmatrix}
\begin{pmatrix}
  t\\\theta
 \end{pmatrix},
\end{equation}
so that
\begin{equation}\label{eq:nearextafterGL}
 \begin{pmatrix}
  \partial_{t'}\\\partial_{\theta'}
 \end{pmatrix}=
\begin{pmatrix}
b&0\\2\pi\ell \,T_R&2\pi\ell\, T_L
\end{pmatrix}
\begin{pmatrix}
 \partial_\tau\\\partial_u
\end{pmatrix}~.
\end{equation}
Here we have included an arbitrary $b>0$ constant, which is equivalent to $b=1$ by diffeomorphism invariance.

The near-extremal limit is now well-defined in coordinates $t'$ and $\theta'$. By simply setting $T_R=0$ in \eqref{eq:nearextafterGL} we get
\begin{align*}
  \partial_{t'}&= b \, \partial_\tau
  \\\partial_{\theta'}&= 2\pi\ell\, T_L \partial_u~.
\end{align*}
This identification gives the self-dual geometry with $\alpha= 2\pi\ell\, T_L$ in accelerating coordinates. The identification with \eqref{eq:metricb} is made by $\phi = \theta'=u/\alpha$ and $\tau=b\, t'$. 

It is useful to describe the limit explicitly in coordinates. For this, we reuse the accelerating coordinate $x$, which is related to $r$ via \eqref{eq:BHrWRTx}. Recall that $x$ is given linearly by $g(\partial_\tau,\partial_u)$ and so it remains invariant under the transformation \eqref{eq:selflimitAB}. We also use the coordinates $(\theta',t')$ from \eqref{eq:selflimitAB}. Altogether we have
\begin{equation}\label{eq:nenhLimitCoord}\begin{aligned}
r&=\frac{r_+-r_-}{2}x + \frac{r_++r_-}{2}\\
t&=-\frac{2\nu}{\nu^2+3}b\frac{T_L}{T_R}t'\\
\theta&=\phi+ \frac{b}{2\pi\ell \, T_R} t'.
\end{aligned}\end{equation}
The ADM metric at fixed $T_L$ and $T_R>0$ in $(t',x,\phi)$ coordinates is
\begin{multline}\label{eq:metricInxtprime}
 g=-\frac{\ell^2}{\nu^2+3}b^2(x^2-1)\left(\frac{4\pi\nu\ell \, T_L}{ R(r(x)) (\nu^2+3)}\right)^2 \,dt'^2 \\+\frac{\ell^2}{\nu^2+3}\frac{dx^2}{x^2-1}
 +\ell^2 R^2(r(x))\left( d\phi + N_{t'}(r(x)) \,dt'\right)^2,
\end{multline}
with
\begin{equation}\label{eq:Ntprime}
 N_{t'}(r) = \frac{b}{2\pi\ell \, T_R}\left(1-\frac{2\nu}{\nu^2+3} 2\pi\ell\, T_L \frac{2\nu r-\sqrt{r_- r_+ (\nu^2+3)}}{2 R^2(r)}\right)~.
\end{equation}

Note that in the limit $r_+\rightarrow r_-$, $r(x)\rightarrow \tfrac{r_++r_-}{2}$. We also have that
\begin{equation*}
 R^2(\frac{r_++r_-}{2})= \frac{ (\nu^2+3)(2\pi \ell \, T_R)^2+(4\pi\nu\ell \, T_L)^2}{(\nu^2+3)^2}\stackrel{r_+\rightarrow r_-}{\longrightarrow} \left(\frac{4\pi\nu\ell \, T_L}{\nu^2+3}\right)^2~.
\end{equation*}
By using the above, and the equations for $T_L$ and $T_R$ in \eqref{eq:TR} and \eqref{eq:TL}, one sees that the term in parentheses in \eqref{eq:Ntprime} is zero as $r_+\rightarrow r_-$. Therefore, it cancels the pole in $T_R$. In order to find the limit we expand the function
\begin{equation*}
 f\left(\frac{r_+-r_-}{2}x + \frac{r_++r_-}{2};r_+,r_-\right) = \frac{2\nu r-\sqrt{r_- r_+ (\nu^2+3)}}{2 R^2(r)}~,
\end{equation*}
 which is symmetric in its last two arguments, in powers of $L$, with $r_\pm=r_{e}\pm L$ and keeping $r_e$ and $x$ fixed:
\begin{multline*}
  f(L\,x+r_e;r_e+L,r_--L)=f(r_e;r_e,r_e)+f^{(1,0,0)}(r_e;r_e,r_e)L\, x \\+ f^{(0,1,0)}(r_e;r_e,r_e)L - f^{(0,0,1)}(r_e;r_e,r_e)L+\order(L^2)~\\ \\
= \partial_r\left. \left( \frac{2\nu r-\sqrt{r_- r_+ (\nu^2+3)}}{2 R^2(r)} \right)\right|_{r=r_-=r_+}\!\!\!\!\!\! \cdot\frac{r_+-r_-}{2}x+\order({T_R}^2)~.
\end{multline*}
After some algebra, we find
\begin{equation*}
 \lim_{r\rightarrow r_+}N_{t'}(x) = \frac{1}{R_-}\frac{2\nu}{\nu^2+3}b\, x~.
\end{equation*}
With $R_-=4\pi\nu\ell \, T_L/(\nu^2+3)$, we confirm that the metric \eqref{eq:metricInxtprime} becomes at $r_+\rightarrow r_-$ the self-dual solution with $\alpha=2\pi\ell\, T_L$ and $t'=b\,\tau$.


Observe that the bifurcate nature of the horizon is inherited in accelerating coordinates. Although this is not a metric limit, in the sense that we have not fixed a black hole geometry, we intuitively understand \eqref{eq:nenhLimitCoord} as zooming in close to the outer horizon of non-extremal black holes with $T_R\approx0$. Finally note that, in taking $T_R\rightarrow 0$, we can keep $T_L$ or some other combination of $T_L$ and $T_R$ fixed. The interpretation of the near-horizon limit in the context of the 4d Planck scale limit $L_p\rightarrow 0$ for Reissner-Nordstr\"om black holes  has been discussed in \cite{Spradlin:1999bn,Maldacena:1998uz}, see also \cite{Maldacena:1998bw}.

\subsection{Extremal limit}\label{sec:extrlim}
In the ADM form one can reach the extremal black holes by setting $r_+=r_-$ in \eqref{eq:BlackHoleADM}. We can describe this by combining the limit $T_R\rightarrow 0$ with an $\text{SL}(2,\mathbb{R})_R$ transformation,
\begin{equation}\label{eq:partialthetaAB}
 \partial_{\theta'} =2\pi \ell \, T_R \left( A \, r_2 \pm B \, r_0\right) +2\pi\ell\, T_L \, l_2,
\end{equation}
where we set
\begin{equation}\label{eq:SL2RforTRLimit}\begin{aligned}
 A &=\frac{1}{\ell  \, T_R}~,&B&=\sqrt{\left({\frac{1}{\ell \, T_R}}\right)^{2}-1}~.
\end{aligned}\end{equation}
In the limit $T_R\rightarrow 0$ we have
\begin{align*}
  \partial_t &=\frac{\nu^2+3}{2\nu}\partial_u\\
  \partial_{\theta'} &=2\pi ( r_2 \pm r_0)+2\pi\ell \, T_L l_2,
\end{align*}
which describe precisely the extremal black holes. Here we do not need to use a $\mathrm{GL}(2,\RR)$ transformation.

\begin{figure}
\begin{center}
 \input{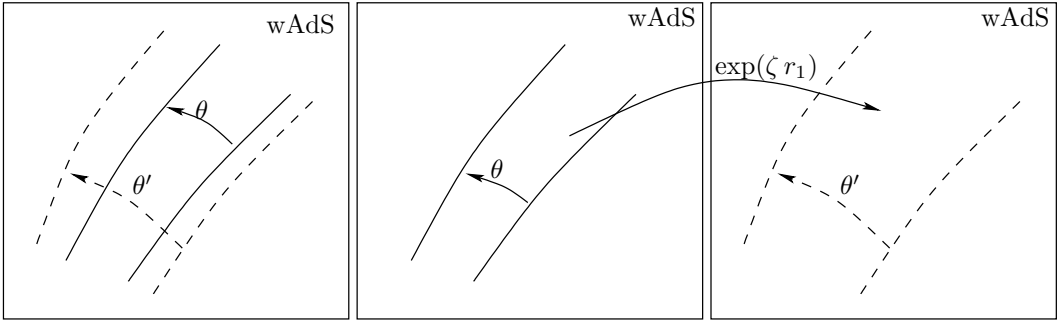}
\caption{The field $r_1$ is not an isometry of the black hole metric, since it does not preserve the identification. However, the mapped region is by definition isometric to the black hole.}\label{fig:notlocaliso}
\end{center}
\end{figure}

We claim that this limit is equivalent to setting $r_- = r_+$ in the ADM form. Indeed, in section \ref{sec:quotients} we only considered the case when $\partial_\theta$ is a linear combination of $r_2$ and $l_2$. Since $e^{\zeta r_1}$ is invertible, the identification along $\partial_\theta$ is equivalent to the identification along $\partial_{\theta'}$:
$$ e^{2\pi\partial_\theta}p\sim p \Longleftrightarrow e^{2\pi\partial_{\theta'}} e^{\zeta r_1}p\sim e^{\zeta r_1} p$$
for every point $p$ in warped AdS. We can define coordinates $(r',t',\theta')$ on the mapped region by using the $(r,t,\theta)$ coordinates of \S\ref{sec:quotients}, with $r'=r$, $t'=t$, $\theta'=\theta$, see figure \ref{fig:notlocaliso}. 

By using the invariant description of the identification vector, it is obvious that in sending $T_R\rightarrow0$, and keeping $T_L$ finite, non-extremal black holes can either limit to the near-extremal geometry with $\alpha= 2\pi\ell \, T_L$, or the extremal black hole with the same $T_L$. That is, we can either try to keep the term in $\partial_\theta$ that is multiplied by $T_R$ (the extremal limit) or not (the near-extremal limit).

\subsection{Near-horizon geometry, again}\label{sec:nhliminv}

We are now able to describe the near-horizon geometry of the extremal black holes, which was given in \S\ref{sec:nhGeomCoord}, in an invariant way. Let us accordingly switch to Poincar\'e coordinates $(x,\tau,u)$. From \eqref{eq:quotientthetaTLTR} and by using an $\mathrm{SL}(2,\RR)_R$ rotation generated by $r_1$, the identification vector is
\[\partial_\theta=2\pi\,L\,(r_2+ r_0) + 2 \pi\ell \, T_L \, l_2~\text{ with $L>0$}.  \] 
It is also necessary to use a matrix transformation as in \S\ref{sec:SDLimit}, so that $\partial_t$ is not collinear with $\partial_\theta$ in the limit $L\rightarrow0$. We use a matrix transformation identical in form to \eqref{eq:selflimitAB}, but replace $T_R$ with $L$. In the limit $L\rightarrow 0$, we obtain the self-dual solution in Poincar\'e coordinates, with $\alpha=2\pi\ell \, T_L $:
\begin{align*}
 \partial_t &= b \, \partial_\tau\\
\partial_\theta &= 2\pi\ell\, T_L \partial_u~.
\end{align*}

One can use coordinates to describe the above limit. In fact, the coordinate transformation follows closely \S\ref{sec:SDLimit}, with some minor changes. In \eqref{eq:nenhLimitCoord}, the first equation should be replaced with $x=L\,(r-r_-)$, and $T_R$ should be replaced with $L$ in the other two equations. The metric in $(r',t',\phi')$ coordinates, \eqref{eq:metricInxtprime}, becomes
\begin{multline*}
 g=-\frac{\ell^2}{\nu^2+3}b^2\,x^2\,\left(\frac{4\pi\nu\ell \, T_L}{ R(r) (\nu^2+3)}\right)^2 \,dt'^2 +\frac{\ell^2}{\nu^2+3}\frac{dx^2}{x^2}
 +\ell^2 R^2(r)\left( d\phi + N_{t'}(r) \,dt'\right)^2~,
\end{multline*}
and, in the limit $L\rightarrow 0$, the metric limits to the self-dual geometry in Poincar\'e coordinates, with $\alpha=2\pi\ell\,T_L$ and $t'=b\,\tau$.

It might seem surprising that this is the same limit as in \S\ref{sec:nhGeomCoord}. Observe however that $\partial_{t'}-\partial_{\phi}$ is proportional to the Killing vector that is null on the horizon. In using the matrix transformation we are rescaling the ingoing coordinate as before. The radial coordinate is then rescaled appropriately so that the limit is finite.

\subsection{Vacuum limit}\label{sec:vaclim}
We finally consider the limit $T_R,T_L\rightarrow 0$ with the ratio $T_L/T_R$ kept constant. This is equivalent to keeping a fixed ratio $r_+/r_-$ and sending $r_-\rightarrow 0$. In \cite{Andy08} this limit was called the vacuum solution. In order to keep $\partial_\theta$ finite, we use the $\mathrm{SL}(2,\RR)_R$ transformation in \eqref{eq:partialthetaAB}, with the same parameters \eqref{eq:SL2RforTRLimit}, so that in the limit $T_R\rightarrow 0$ we obtain
\begin{equation}\label{eq:VacVectors}\begin{aligned}
  \partial_t &=\frac{\nu^2+3}{2\nu}\partial_u\\
  \partial_\theta &=2\pi ( r_2 \pm r_0)~.
\end{aligned}\end{equation}
Here we did not need the $\mathrm{GL}(2,\RR)$ transformation. Observe that the Killing vectors $\partial_t$ and $\partial_\theta$ do not depend on $r_+/r_-$. The limit is thus universal and there is no remnant of the ratio $T_L/T_R$. The geometry we obtain by \eqref{eq:VacVectors} is warped AdS$_3$ in Poincar\'e coordinates with a periodic identification of the proper time. That is, the metric \eqref{eq:PoincareMetric}
\begin{equation}
g_{\ell,\nu}=\frac{\ell^2}{\nu^2+3}\left(-x^2 d\tau^2+\frac{dx^2}{x^2}+\frac{4\nu^2}{\nu^2+3}\left(du+x\, d\tau\right)^2\right),
\end{equation} 
with $\tau=\theta$ and $ u= \frac{\nu^2+3}{2\nu} t $, hence $\tau\sim \tau+2\pi$ and $x,u\in \RR$. 

We can see the same result by using coordinates. As in the extremal limit, we use the metric in ADM form, and we send the parameters $r_-$ and $r_+$ to zero keeping $r_+/r_-$ fixed. The metric becomes
\begin{multline}\label{eq:vacuummetric}
 \lim_{\substack{r_-\rightarrow 0\\r_+\rightarrow 0}}g
= - \ell^2 \frac{\nu^2+3}{3(\nu^2-1)} dt^2 + \frac{\ell^2}{\nu^2+3}\frac{dr^2}{r^2}+\ell^2\frac{3(\nu^2-1)}{4}r^2\left(d\theta+\frac{4\nu}{3(\nu^2-1)}\frac{1}{r}dt\right)^2\\
\!\!=-\ell^2\frac{\nu^2+3}{4}r^2 d\theta^2 +\frac{\ell^2}{\nu^2+3}\frac{dr^2}{r^2} + \frac{4\nu^2\ell^2}{(\nu^2+3)^2}\left(\frac{\nu^2+3}{2\nu} dt +\frac{\nu^2+3}{2} r \,d\theta\right)^2~.
\end{multline}
The identification with Poincar\'e coordinates can be made with $x=\frac{\nu^2+3}{2} r$.

The limit corresponds to sending $M_{\text{\scriptsize{ADT}}}$ and $J_{\text{\scriptsize{ADT}}}$ to zero while keeping the Hawking temperature fixed, see appendix C. One can also interpret this as a limit to the far-away region. That is, the metric in \eqref{eq:vacuummetric} corresponds to keeping the leading order components of the black hole metric when $r\gg r_+$. This agrees with the fixed components of the asymptotic boundary conditions of \cite{Compere:2009zj}, which we recall in appendix C.

\section{Discussion}\label{sec:Discussion}

In this paper we explored the geometry of black holes in spacelike stretched warped AdS. We elaborated on the construction of warped $\AdS_3$ from first principles, described suitable coordinates, and performed the quotient construction. We focused on the case when causal singularities do exist and are hidden behind a Killing horizon. The geometries are ideal, in the sense that they can be continued to regions that contain new singularities and new asymptotic regions. We found the causal structure and showed that the geometries fall into three classes that resemble the causal structure of the Reissner-Nordstr\"om black hole.

We pointed out two features that are usually suppressed in the literature. The first is that the black hole metric parametrized by $r_+$ and $r_-$  presents a redundancy, in that for a certain region two sets of parameters $(r_+,r_-)$ describe the same geometry. The second is that, the ratio of the left to right temperature is bounded from below, if the geometry is to describe a causal singularity that is hidden behind Killing horizons. In \cite{Compere:2009zj} care was taken to consistently define an asymptotically Killing algebra \cite{Brown:1986nw} that contains a centrally extended Virasoro algebra with generators $\mathscr{L}_m$, so that $\mathscr{L}_0$ has positive spectrum and a central extension that matches the AdS/CFT expectation \cite{Cardy:1986ie}. The bound on the ratio of temperatures $T_L/T_R$ would then imply an upper bound on $\mathscr{L}_0$.

We also described various spacetime limits that one can take in the black hole phase space. We do this by studying the behavior of the identification vector $\partial_\theta$ for different significant limits of the invariants $T_R$, $T_L$ and their ratio. We chose this exposition for the clarity of the geometric interpretation of the limits, and also to avoid the ambiguities that could come from a coordinate description. In this description, it is easy to see that the possible limits using this method are again quotients of warped AdS. Furthermore, the spacetime limits inherit suitable coordinates that are not global. In particular, we get the self-dual solution in accelerating or Poincar\'e coordinates, and warped AdS$_3$ in Poincar\'e coordinates under a proper time identification.

The spacelike stretched black holes are a subset of the general black holes of cosmological Einstein-Maxwell theory with gravitational and gauge Chern-Simons couplings, which were presented in \cite{Moussa:2008sj}, with $\mu_E/\mu_G=2/3$ and $\beta^2=(\nu^3+3)/(4\nu^2)$. There, the causal structure of the general black holes was also first reported. We have here presented an \emph{explicit} Kruskal extension, which underlies the Penrose diagram of the maximal extension. Our derivation focuses on the metric $g_2$ that is defined on the two-dimensional quotient space of a black hole by the global isometry $\partial_\theta$. One can successively remove detail from our presentation but retain the reduction on $g_2$, since this captures the essential causal relations of the 3d spacetime.

Also in \cite{Moussa:2008sj}, local coordinate transformations were given that relate the various black holes, the self-dual solution, and the vacuum. Here, we only write the local coordinate transformations between the black holes, or the self-dual solution, and spacelike warped AdS, which precisely define the first as discrete quotients of the latter. Furthermore, the vacuum and self-dual solution are obtained here as \emph{limits} of the black holes. This was done invariantly using the identification vector, but also through well-defined coordinate transformations. We note this comparison so as to highlight the structure of this work. Let us also remark that the limits we consider are classical, that is $\nu$, $G$ and $\ell$ are kept fixed. This does not allow us to obtain, for instance, the black holes with vanishing cosmological constant \cite{Moussa:2003fc}.

Let us also compare to the construction to the Banados-Teitelboim-Zanelli black holes of Einstein gravity with a negative cosmological constant. The BTZ black holes are necessarily asymptotically conformally flat. Therefore, the conformal boundary is always timelike. The causal diagrams of the BTZ black holes fall into two classes, depending on whether the geometry is extremal or not~\cite{Banados:1992gq,Banados:1992wn}. This is different to the warped AdS$_3$ case, which is not conformally flat.

One motivation for this work was to find a non-extremal spacetime limit where the acceleration coordinate $\partial_\tau$ would explicitly depend on a parameter $b$. This would imply that the limit inherits two parameters rather than the one in $u=2\pi\ell\, T_L\,\phi$. Then one could approximate the chiral thermal Green functions of the near-extremal black holes with those computed in the self-dual warped AdS$_3$ space in accelerating coordinates, see \cite{Chen:2010qm,Li:2010zr}. It is for this reason that we introduced the constant $b$ in \eqref{eq:metricInxtprime}. By diffeomorphism invariance though, we can set this constant equal to 1. We speculated on whether a suitable set of asymptotic conditions can break this freedom. 

Topological massive gravity is expected to have a rich spectrum and we believe that the solution space will present new insight in the AdS/CFT correspondence. Understanding better the relation of TMG and its solutions to four-dimensional reality is another direction we look forward to.

\section*{Acknowledgment}
The authors would like to thank Gerard Clement for useful advice and discussions.

\appendix

\section{Hyperbolic fibration}

In section \ref{sec:geometry} we observed that the warped metric in waped coordinates
\[g_{\ell,\nu}=\frac{\ell^2}{\nu^2+3}\left(-\cosh^2\sigma \, d{\tilde{t}}^2+d\sigma^2+\frac{4\nu^2}{\nu^2+3}\left(d{\tilde{u}}+\sinh\sigma\, d{\tilde{t}}\right)^2\right)~\]
is compatible with a double cover of $\mathrm{SL}(2,\RR)$. That is, 
 the time identification on $\mathrm{SL}(2,\RR)$ is $\tilde{t}\sim\tilde{t}+4\pi$, whereas the base space describes a two-dimensional quadric with $\tilde{t}\sim\tilde{t}+2\pi$. 

\begin{figure}
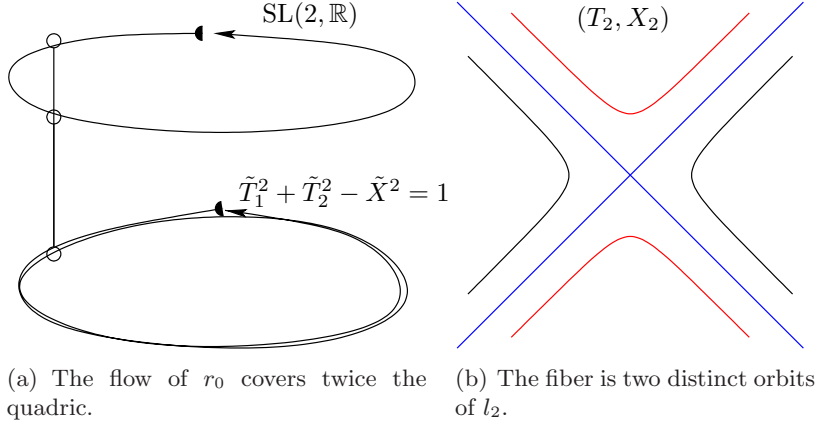

\begin{center}
 \subfigure[{The flow of $r_0$ covers twice the quadric.}]{{\input{figtimeflow.pspdftex}}\label{fig:timeflow}
}\quad
\subfigure[{The fiber is two distinct orbits of $l_2$.}]{{\input{figT2X2plane.pspdftex}}\label{fig:T2X2plane}
}\caption{Hyperbolic fibration}
\end{center}
\end{figure}

Here we elaborate on this without using warped coordinates. The vector field $l_2$ in \eqref{eq:tauvectors} defines a non-trivial real-line fibration of the quadric
 \begin{equation}\label{eq:3quad}  T_1^2+T_2^2-X_1^2-X_2^2=1 \end{equation}
 over the quadric
\begin{equation}\label{eq:AdS2Quad}
\tilde{T}_1^2+\tilde{T}_2^2-\tilde{X}^2=1.
\end{equation}
Explicitly, the projection
\begin{equation}\label{eq:ProjDQuad}\begin{aligned}
 \tilde{T}_1-\tilde{X} &= 2(X_1+T_1)(X_2-T_2)\\
 \tilde{T}_1+\tilde{X} &= 2(X_1-T_1)(X_2+T_2)\\
 \tilde{T}_2 & = 2 (T_2^2-X_2^2)-1~
\end{aligned}\end{equation}
is invariant under $l_2$ and satisfies \eqref{eq:AdS2Quad}. Conversely, for every point
$(\tilde{T}_1,\tilde{T}_2,\tilde{X}_1)$ that satisfies \eqref{eq:AdS2Quad}, there are two real-line orbits 
 that satisfy \eqref{eq:3quad} and \eqref{eq:ProjDQuad}. Indeed, \eqref{eq:ProjDQuad} can be solved depending
on the value of $\tilde{T}_2\,$: if $\tilde{T}_2<-1$ the solutions
will cross $T_2=0$ and the two orbits are distinguished by the sign of
$X_2$; similarly, if $\tilde{T}_2>-1$ the same happens, but with $T_2$ and $X_2$ exchanged; if  $\tilde{T}_2=-1$ the two orbits are given by $T_2=\pm X_2$. This is shown in figure  \ref{fig:T2X2plane}. Furthermore, the two orbits are connected by the action of $r_0$. On the base space, $r_0$ induces a
rotation of period $2\pi$:
\begin{align*}
 \lie_{r_0} (\tilde{T}_2)&=\tilde{T}_1\\\lie_{r_0} (\tilde{T}_1)&=-\tilde{T}_2\\
 \lie_{r_0} (\tilde{X})&=0,
\end{align*}
while, from \eqref{eq:paramTTXX}, in
$\text{SL}(2,\RR)$  it has a period of $4\pi$. This is schematically depicted in figure \ref{fig:timeflow}.  

Let us note that the hyperbolic fibration of $\mathrm
{SL}(2,\RR)$ is different to the Hopf fibration of the three-sphere, in that the latter covers the two-sphere once. If two complex numbers $z_1$, $z_2$ are used to describe the three-sphere as $|z_1|^2+|z_2|^2=1$, then the projection of the Hopf fibration is $\pi(z_1,z_2)=(2 z_1
  z_2^\ast,|z_1|^2-|z_2|^2)\in S^2$. For every point in $S^2$ there is then precisely one orbit in $S^3$ given by the action $(z_1,z_2)\mapsto(e^{i\theta}z_1,e^{i\theta}z_2)$. 

\section{Diffeomorphisms}
In writing the warped metric in accelerating coordinates we did not need to use an explicit diffeomorphism. Indeed, using the definition $x=\tfrac{(\nu^2+3)^2}{4\nu^2\ell^2}g_{\ell,\nu}(\partial_u,\partial_\tau)=\cosh\sigma\sin \tilde t$, we were able to calculate the metric components $g_{\tau\tau}=g(r_2,r_2)$, $g_{\tau u}=g(r_2,l_2)$, $g_{uu}=g(l_2,l_2)$ and $g^{xx}=g^{-1}(dx,dx)$ in terms of $x$. Let us present here an explicit diffeomorphism of the accelerating patch.
  
The region $x>1$ with metric \eqref{eq:metricb} isometrically embeds in warped AdS$_3$ under
\begin{equation}\begin{aligned}\label{eq:AdS2trans}
  \sinh\sigma&=\sqrt{x^2-1}\cosh\tau\\
 \cot \tilde{t}&=-\frac{\sqrt{x^2-1}}{x}\sinh\tau\\
 \tilde{u}&=u+\tanh^{-1}(\frac{\tanh\tau}{x})~.
\end{aligned}\end{equation}
This covers $\tilde{u}\in\RR$, $\sigma>0$, and $\tilde{t} \in (0,\pi)$ with $\cosh \sigma \sin {\tilde{t}}>1$. The inverse of \eqref{eq:AdS2trans} is
\begin{equation*}
 \begin{aligned}
  x&=\cosh\sigma\sin \tilde{t}\\
  \tanh\tau&=-\coth\sigma\cos \tilde{t}\\
u&=\tilde{u}+\tanh^{-1}\frac{\cot \tilde{t}}{\sinh\sigma}~,
 \end{aligned}
\end{equation*}
which is well-defined for $\sigma>0$, $\tilde{t}\in(0,\pi)$ and
\[ \left|\frac{\cot \tilde{t}}{\sinh\sigma}\right|<1 \Leftrightarrow |\cosh\sigma \sin \tilde{t}| >1 ~.\]
Similarly the region $|x|<1$ can be embedded with
\begin{equation}
 \begin{aligned}
  \label{eq:AdS2trans2}
 \sinh\sigma&=\sqrt{1-x^2}\sinh\tau\\
 \tan \tilde{t} &=-\frac{x}{\sqrt{1-x^2}}\frac{1}{\cosh\tau}\\
 \tilde{u}&=u+\tanh^{-1}({x\tanh\tau})~,
 \end{aligned}
\end{equation}
whose inverse is
\begin{equation*}
 \begin{aligned}
  x&=\cosh\sigma\sin \tilde{t}\\
 \tanh \tau&=-\frac{\tanh\sigma}{\cos \tilde{t}}\\
 u&=\tilde{u}+\tanh^{-1}({\sinh\sigma\tan \tilde{t}})~.
 \end{aligned}
\end{equation*}
Here we cover $\sigma\in\RR$, $\tilde{u}\in\RR$ and 
\[\left|{\sinh\sigma\tan \tilde{t}}\right|<1 \Leftrightarrow | \cosh\sigma\sin \tilde{t}| <1 ~.\]

Similarly, let us give an explicit diffeomorphism between the Poincar\'e  and warped coordinates and some detail on how we derived it. First we show that in Poincar\'e coordinates, $r_1= x\partial_x-\tau\partial_\tau$ follows from
\begin{equation}\label{eq:r1onPoincx}
 {r_1} (x)  = \frac{(\nu^2+3)^2}{4\nu^2 \ell^2}\lie_{r_1}(g_{\ell,\nu}(\partial_u,\partial_\tau))=\frac{(\nu^2+3)^2}{4\nu^2 \ell^2}g_{\ell,\nu}(\partial_u,[r_1,\partial_\tau])=x
\end{equation}
\begin{align}
[r_1,\partial_\tau]= \partial_\tau&\Rightarrow \partial_\tau({r_1}( \tau))=-1\text{ and }\partial_\tau({r_1}( u))=0 \\
[r_1,\partial_u]=0&\Rightarrow \partial_u({r_1}( u)) = 0\text{ and }\partial_u({r_1}(\tau)) =0 ~,
\end{align}
where we use that $r_1$ is Killing and the commutation relations. Indeed, note how rescaling $x\mapsto e^{\zeta} x$ and $\tau\mapsto e^{-\zeta}\tau$ is an isometry of the metric in Poincar\'e coordinates. Using this and the relation 
\begin{equation}\label{eq:PD1}
x=\tfrac{(\nu^2+3)^2}{4\nu^2
  \ell^2}g_{\ell,\nu}(\partial_u,\partial_\tau) =
\sinh\sigma+\sin {\tilde{t}}\cosh\sigma~,
\end{equation}
we compute
\[\partial_t x= - \lie_{r_0} x =\tfrac{(\nu^2+3)^2}{4\nu^2 \ell^2}g_{\ell,\nu}(\partial_u,r_1)~,\]
or equivalently
\begin{equation}\label{eq:PD2}
x\tau=-\cos {\tilde{t}} \cosh\sigma~.
\end{equation}
Finally, integrating $\sinh\sigma d\tilde{t}-x d\tau$ gives us
\begin{equation}\label{eq:PD3}
 u=\tilde{u}+\ln\left(\pm\frac{\cosh{\sigma/2}\cos{t/2}+\sinh{\sigma/2}\sin{t/2}}{\cosh{\sigma/2}\sin{t/2}+\sinh{\sigma/2}\cos{t/2}}\right)~.
\end{equation}
The explicit diffeomorphism is given by \eqref{eq:PD1}, \eqref{eq:PD2} and \eqref{eq:PD3}.  We confirm that they are well-defined and $x\lessgtr0$ is equivalent to
\[\frac{\cosh{\sigma/2}\cos{t/2}+\sinh{\sigma/2}\sin{t/2}}{\cosh{\sigma/2}\sin{t/2}+\sinh{\sigma/2}\cos{t/2}}\lessgtr0~.\]

The quotient construction of section \ref{sec:quotients} uses a linear relation between the black hole coordinates $(t,\theta)$ and the accelearting or Poincar\'e coordinates. Combining the linear relation with the above diffeomorphisms, one can write the diffeomorphism of the black hole coordinates with respect to the warped coordinates that define the quotient:
 \begin{align*}
   \tilde{t}& = \tan^{-1} \left( \frac{2\sqrt{(r-r_+)(r-r_-)}}{2\,r-r_+-r_+} \sinh \left(\frac{1}{4}(r_+-r_-)(\nu^2+3)\theta \right) \right)\\
   \sigma&=\sinh^{-1}\left( \frac{2\sqrt{(r-r_+)(r-r_-)}}{r_+-r_+}\cosh(\frac{1}{4}(r_+-r_-)(\nu^2+3)\theta) \right)\\
   \tilde{u}&=\frac{\nu^2+3}{4\nu}\left( 2 \, t + \left( \nu(r_++r_-)-\sqrt{r_+ r_- (\nu^2+3)}\right)\theta \right)\\
   &\quad+\coth^{-1}\left(  \frac{2r-r_+-r_-}{r_+-r_+}\coth\left( \frac{1}{4}(r_+-r_-)(\nu^2+3)\theta\right) \right)~.                   
 \end{align*}
This is essentially the transformation in eqs.(5.3)-(5.5) of \cite{Andy08}, although here they are defined in $r>r_+$ as opposed to $r_-<r<r_+$. Note that we have translated $\tilde{t}\mapsto\tilde{t}+\tfrac{\pi}{2}$ with respect to \eqref{eq:AdS2trans}.

\section{Thermodynamics}
We would like to recall here the thermodynamic quantities that were computed for the spacelike warped black holes in \cite{Andy08}. It is also noteworthy to translate the bound on the ratio of temperatures and the region $r_+/r_-< \lambda_f$ into conventions used in the literature. However, let us briefly comment on the ADM form. A general stationary, axisymmetric, asymptotically-flat black hole uniquely normalizes the Killing vector
\begin{equation*}
 \xi = \partial_t - \Omega \partial_\theta~
\end{equation*}
that is null on its horizon, by using the asymptotically defined $t$ and $\theta$. Quantities like the surface gravity $\kappa_0=2 \pi T_H$ on its horizon $\mathscr{H}$ are unambiguously defined, e.g.
\begin{equation*}
 \nabla_\xi \xi \stackrel{\mathscr{H}}{=} \kappa_0 \xi ~.
\end{equation*}
Were we to use a different time and angle
\begin{equation}\label{eq:otherADM}\begin{aligned}
 t' &= \Lambda \, t\\
 \theta' &= \theta + b\, t~,
\end{aligned}\end{equation}
the Hawking temperature, angular velocity $\Omega$, and ADT charges~\cite{Abbott:1982jh,Bouchareb:2007yx,Deser:2002rt,Deser:2002jk,Deser:2003vh}, here the mass $M_{\text{\scriptsize{ADT}}}$ and angular momentum $J_{\text{\scriptsize{ADT}}}$, would transform as
\begin{align*}
 T_H' &= \frac{1}{\Lambda} T_H\\
 \Omega' &= \frac{\Omega+b}{\Lambda}\\
 \delta M_{\text{\scriptsize{ADT}}}'  &= \frac{1}{\Lambda}\delta M_{\text{\scriptsize{ADT}}} - \frac{b}{\Lambda}\delta J_{\text{\scriptsize{ADT}}}\\
 \delta J_{\text{\scriptsize{ADT}}}'&=\delta J_{\text{\scriptsize{ADT}}}~.
\end{align*}
On the other hand, the entropy variation in the first law, $\delta S = \frac{1}{T_H}(\delta M_{\text{\scriptsize{ADT}}} - \Omega \delta J_{\text{\scriptsize{ADT}}})$, is seen to be invariant under \eqref{eq:otherADM}. The Wald formula for the entropy \cite{Iyer:1994ys} as applied for TMG in \cite{Tachikawa:2006sz} (see also \cite[\S4.2]{Bouchareb:2007yx}) depends on the asymptotic \emph{orthonormal} frame and its spin connection, and therefore is indeed invariant under \eqref{eq:otherADM}. 

We normalize the thermodynamic quantities with respect to the frame where 
\begin{equation*}
g(\partial_t,\partial_t)=\ell^2~.
\end{equation*}
This is compatible to the asymptotically warped $\AdS_3$ conditions in \cite{Compere:2009zj}. In particular, it fixes both $t$ and $\theta$ coordinates as in the ADM form \eqref{eq:BlackHoleADM}. From \cite{Andy08}, we have
 \begin{align*}
  T_H &=
  \frac{\nu^2+3}{4\pi\ell\nu}\frac{T_R}{T_L+T_R}\\
  \Omega &= -\frac{\nu^2+3}{4\pi\nu} \frac{1}{T_R+T_L}\\
  M_{\text{\scriptsize{ADT}}}&= 
  \frac{\pi}{3 G}\ell \, T_L\\
  J_{\text{\scriptsize{ADT}}} &= 
  \frac{\nu \ell }{3 (\nu^2+3)G}\left( \left( 2\pi\ell\, T_L \right)^2 - \frac{5\nu^2+3}{4 \nu^2}\left(2\pi\ell \, T_R\right)^2\right)\\
  S &= \frac{\pi^2 \ell}{3}\left( \frac{5\nu^2+3}{\nu(\nu^2+3)G}\ell \, T_R + \frac{4 \nu}{(\nu^2+3)G}\ell \, T_L \right)~.
 \end{align*}
The CFT correspondence conjecture in \cite{Andy08} allows one to write the entropy in the form of Cardy's formula \cite{Cardy:1986ie} with left/right central extension charges $c_R= \frac{5\nu^2+3}{\nu(\nu^2+3)G}\ell$ and $c_L=\frac{4 \nu}{(\nu^2+3)G}\ell$. 
The bound in \eqref{eq:lambdaf} and $T_R\geq 0$ become, respectively, the left-hand side and right-hand side of
\begin{equation*}
 -\frac{8\nu\ell G}{\nu^2-1}M_{\text{\scriptsize{ADT}}}^2 \leq J_{\text{\scriptsize{ADT}}} \leq \frac{12\nu\ell G}{\nu^2+3}M_{\text{\scriptsize{ADT}}}^2~.
\end{equation*}

There is yet another form of the black hole metrics\footnote{the black hole metric was first written in \cite{Bouchareb:2007yx}.} that is given in
\cite{Compere:2008cv}, \cite{Compere:2009zj} and \cite{Moussa:2008sj}. The metric in \cite{Compere:2008cv} with parameters $(\nu',J',a',L')$ is related to the one in \cite{Compere:2009zj}, which we write here
\begin{multline}\label{eq:BHCompere}
 ds^2 = 
 dT'^2  +(\frac{3}{\ell^2}(\nu^2-1)R'^2 - \frac{4 \mathfrak{ j}\ell }{\nu}+12 \mathfrak{ m} R' )d\theta^2 
 \\- 4  \frac{\nu}{\ell} R' dT' d\theta +\frac{dR'^2}{\frac{3+\nu^2}{\ell^2}R'^2 - 12 \mathfrak{ m }R' + \frac{4 \mathfrak{ j} \,\ell}{\nu}},
\end{multline}
by $\mathfrak{j}= G J'$, $6\mathfrak{m}=4 G \nu'$, $a'=-\nu/\ell$ and $L'=\sqrt{2\ell/(3-\nu^2)}$. The metric in \eqref{eq:BHCompere} is related to \eqref{eq:BlackHoleADM} under the transformation $R'=\frac{\ell^2}{2}r-\frac{\ell^2}{4\nu}\sqrt{r_+ r_-(\nu^2+3)}$ and  $T'=\ell\, t$ with
\begin{align*}
6 \mathfrak{m} &= \frac{\nu^2+3}{4}\left( r_++r_--\frac{\sqrt{r_+ r_- (\nu^2+3)}}{\nu}\right) = 2\pi\ell \, T_L\\
4\mathfrak{j} &= \frac{5\nu^2+3}{16 \nu}(\nu^2+3) \ell r_- r_+ - \frac{(\nu^2+3)^{\frac{3}{2}}}{8}\ell(r_++r_-)\sqrt{r_-r_+}~.
\end{align*}
The existence of a Killing horizon is given by the vanishing of $g_{R'R'}$ for some value of $R'$. This is a quadratic equation in $R'$ with determinant
\begin{equation*}
 \Delta_{R'}=(12 \mathfrak{m})^2-16\mathfrak{j}\frac{\nu^2+3}{\nu\ell}=\left( \frac{\nu^2+3}{2}(r_+ - r_-) \right)^2=(4\pi\ell \, T_R)^2\geq 0~.
\end{equation*}
On the other hand, there are causal singularities hidden behind a Killing horizon when $g_{\theta\theta}$ vanishes. This is equivalent to $r_+/r_-<\lambda_f$ and gives
\begin{equation*}\mathfrak{j}\geq- 3\mathfrak{m}^2\nu\ell/(\nu^2-1)~.\end{equation*}
For smaller values of $\mathfrak{j}$ for fixed $\mathfrak{m}$ we continue in the region where there are no CTCs.

\bibliographystyle{utphys}
\bibliography{main3}
\end{document}